\documentclass[12pt]{iopart}
\usepackage{iopams}
\usepackage{graphicx}

\begin{document}

\title{A Coherent RC Circuit}
\author{J. Gabelli}
\address{Laboratoire de Physique des Solides (UMR 8502) b\^atiment
510, Universit\'e Paris-Sud 91405 Orsay Cedex, France}
\ead{julien.gabelli@u-psud.fr}

\author{G. F\`eve, J-M. Berroir, B. Pla\c{c}ais }
\address{Laboratoire Pierre Aigrain, Ecole Normale Sup\'erieure, CNRS (UMR 8551), Universit\'e
P. et M. Curie, Universit\'e D. Diderot, 24 rue Lhomond, 75231 Paris Cedex 05, France}

\begin{abstract}
We review the first experiment on dynamic transport in a phase-coherent quantum conductor. In our discussion, we highlight the use of time-dependent transport as a means of gaining insight into charge relaxation on a mesoscopic scale. For this purpose, we studied the ac conductance of a model quantum conductor, \textit{i.e.} the quantum RC circuit. Prior to our experimental work, M. B\"{u}ttiker, H. Thomas and A. Pr\^{e}tre \cite{Phys_Lett_A180_364_Buttiker} first worked on dynamic mesoscopic transport in the 1990s. They predicted that the mesoscopic RC circuit can be described by a quantum capacitance related to the density of states in the capacitor and a constant charge relaxation resistance value equal to half of the resistance quantum $h/2e^2$, when a single mode is transmitted between the capacitance and a reservoir. By applying a microwave excitation to a gate located on top of a coherent submicronic quantum dot that is coupled to a reservoir, we validate this theoretical prediction on the ac conductance of the quantum RC circuit. Our study demonstrates that the ac conductance is directly related to the dwell time of electrons in the capacitor. Thereby, we observed a counterintuitive behavior of a quantum origin: as the transmission of the single conducting mode decreases, the resistance of the quantum RC circuit remains constant while the capacitance oscillates.
\end{abstract}

\maketitle
\begin{table}[h]
\caption{\label{tab:symbols} List of symbols.}
\vspace{0.5cm}
\begin{indented}
\item[]\begin{tabular}{@{}ll}
\br
$e$& elementary charge \\
$h$& Planck constant\\
$\hbar$& reduced Planck constant\\
$k_B$& Boltzmann constant \\
$n_e$& electronic density \\
$\mu_e$& electronic mobility\\
$\tau_{RC}$& RC time\\
$\tau_d$& dwell time \\
$\tau_0$& roundtrips time inside the cavity\\
$C$& geometric capacitance \\
$C_q$& quantum capacitance\\
$C_{\mu}$& electrochemical capacitance capacitance\\
$R_q$& resistance of the charge relaxation \\
$R_K$& von Klitzing constant \\
$R_c$& contact resistance \\
$f$& Fermi-Dirac distribution \\
$\epsilon_F$& Fermi energy\\
$N$& density of states\\
$\mu_{L,R}$& electrochemical potential of the left, right reservoir \\
$\omega$& angular frequency\\
$s_{i,j}$& scattering matrix \\
$\mathcal{T}$& transmission probability \\
$\mathcal{R}$& reflection probability \\
$r$& reflection amplitude\\
$\Delta$& energy level spacing in the cavity\\
$\Delta^{\star}$& renormalized energy level spacing by Coulomb interactions\\
$\hbar \Gamma$& width of the energy level\\
$v_d$& drift velocity\\
$l$& circumference of the catvity \\
$\phi$& accumulated phase in the cavity\\
$g_{dc}$& dc conductance of the 2DEG\\
$g_{ac}$& ac conductance of the 2DEG \\
$G$& conductance\\
$Z$& impedance \\
$V_i$& electric potential of the reservoir\\
$U$& electric potential in the cavity\\
$V_{g}$& gate voltage\\
$X,Y$& in-phase and out-of-phase signals \\
$\varphi$& phase of the signal\\
\br
\end{tabular}
\end{indented}
\end{table}

\newpage

\section{Introduction}

\begin{figure}[h]
\begin{center}
\includegraphics[width=0.45\linewidth]{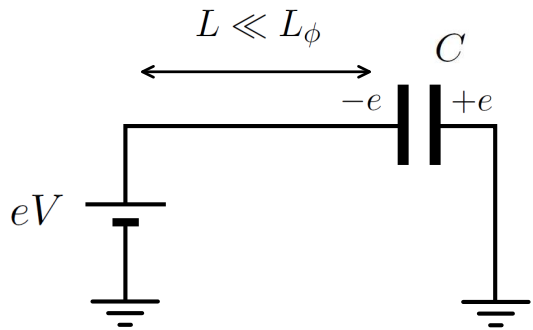}
\end{center}
\caption{\textit{Gedankenexperiment}. An electron is confined to the electrode of a capacitor $C$ by a voltage source: $V=e/C$. The length of the circuit $L$ is assumed to be smaller than the phase coherence length $L_{\phi}$.} \label{fig1}
\end{figure}

In view of recent developments in quantum electronics, the increasing interest in manipulating and measuring a single electronic charge raises the question of whether a quantum limit of the charge relaxation time in an electronic circuit exists. In the classical sense, the charge relaxation corresponds to the exponential decay of charges, while its characteristic time is directly related to both the dissipation in the conductor and the electronic interaction. The charge relaxation of the RC circuit illustrates this correlation. The relaxation time of the charge on a conductor is given by the product $\tau_{RC}=R \times C$, where $C$ is the self-capacitance of the conductor and $R$ is the resistance connecting the conductor's self-capacitance to an electronic reservoir. Thus, the realization of the quantum equivalent of the RC circuit presents a starting point for characterizing the time scales governing the quantum dynamics of electrons. Before describing the quantum RC circuit in detail, let us first imagine a \textit{Gedankenexperiment} involving quantum capacitance. In quantum electronics, a circuit is known to have an electronic wave function that preserves its phase coherence over the whole device. Specifically, the characteristic length of a circuit is smaller than the phase coherence length  ($L \ll L_{\phi}$) \cite{Imry_book,Datta_book,Nazarov_book}. Next let us consider an electron of charge $-e$ , which is confined in the electrode of a capacitor $C$ due to a voltage source at a fixed voltage of  $V=e/C$ (see figure \ref{fig1}). The electronic wave function is delocalized over the electrode and its energy is set to $e^2/C$. If the voltage source suddenly drops to $V=0$, the delocalized electronic wave function reaches an energy uncertainty of $\Delta E=e^2/C$ and according to the Heisenberg's principle, its lifetime is $\tau \sim h/e^2 \, C$. When compared to the RC time, the lifetime leads to a typical resistance of the charge relaxation value of $R_q \sim h/e^2$, even without considering the presence of any dissipative part in the circuit.

Interestingly, this simple \textit{Gedankenexperiment} presents the resistance quantum, which according to the von Klitzing constant, is $R_K=h/e^2$ \cite{ROP_64_1603_Jeckelmann,SemPointcré_Jeckelmann}\footnote{Since 1990, the von Klitzing constant $R_K = 25812.807557(18)\, \Omega$ represents the new definition of the standard of electrical resistance.}. However, the resistance quantum usually applies to a dc transport experiment. A question that arises from this observation is whether $R_K$, which we introduced through a dimensional argument, is exactly the quantum of charge relaxation resistance involved in dynamic transport.

\begin{figure}[h]
\begin{center}
\includegraphics[width=0.6\linewidth]{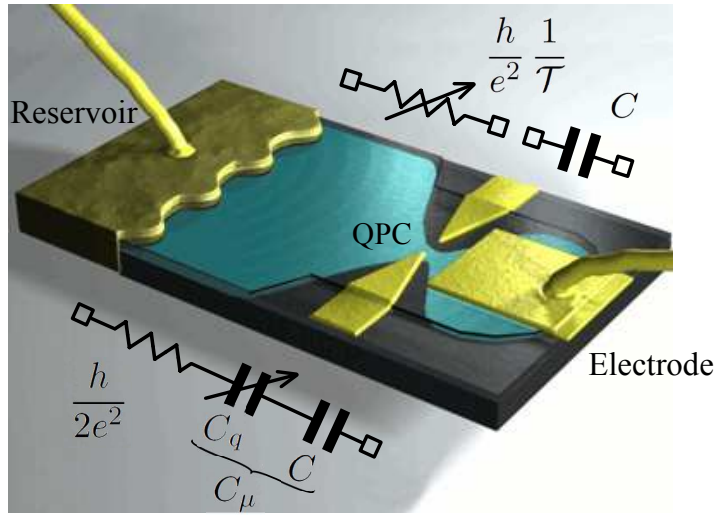}
\end{center}
\caption{The quantum RC circuit was realized using a 2DEG. The capacitor consists of a metallic electrode (gold) on top of a submicrometer 2DEG quantum dot (blue) that defines the second electrode. The resistor is a QPC connecting the dot to a wide 2DEG reservoir (blue), which itself is connected to a metallic contact (dark gold). The QPC controls the number of electronic modes and their transmission. Top: The equivalent circuit of the QPC and the geometric capacitance considered separately. Bottom: The equivalent circuit of the coherent RC circuit. As predicted by the B\"{u}ttiker theory, the relaxation resistance $R_q=h/2e^2$, which enters the equivalent circuit for the coherent conductance, is transmission-independent and equal to half of the resistance quantum. The capacitance $C_{\mu}$ is the serial combination of the quantum and the geometric capacitances ($C_q$ and $C$, respectively). $C_q$ is transmission-dependent and strongly modulated by gate voltages.} \label{fig2}
\end{figure}

$R_K$ usually refers to the integer quantum Hall effect (IQHE), which can be observed at sufficiently low temperature in a two-dimensional electron gas (2DEG) in the presence of a perpendicular magnetic field  \cite{PRL_45_494_Klitzing}. In this specific case, the 2DEG has quantized energy levels called Landau levels, while the Hall resistance exhibits plateaus at quantized values of $R_n=R_K/n$ with $n$ integer. However, the presence of a magnetic field is not necessary for observing such a quantization. Another well-known phenomenon revealing quantized resistance is demonstrated by electron transport through quantum point contacts (QPC), the quantum equivalent of a resistor \cite{PRL_60_848_Wees}. A QPC comprises of a tunable narrow constriction connecting two large conductors. Since the constriction behaves as an electronic waveguide, the dc resistance decreases in quantized steps $R_n$, as the constriction widens. Each step corresponds to an electron-propagating mode in the waveguide \footnote{At zero magnetic field, the resistance of a QPC is actually $R_n=R_K/(2n)$. The factor 2 appears as a result of spin degeneracy. In the present review, all the measurements were performed in the presence of a magnetic field, resulting in the spin degeneracy being lifted.}. Although these two dc experiments appear different, they can be described within the same theory. In this regard, the insights of R. Landauer and M. B\"{u}ttiker provide an exemplary description of electronic transport at sufficiently low temperature by viewing charge transport as a quantum transmission of conducting channels \cite{PRL_31_6207_Buttiker}. With this in mind, the dc conductance is then given at a low temperature level by the so-called Landauer-B\"{u}ttiker formula for a two-contact conductor:

\begin{equation}
G = \frac{e^2}{h} \, \sum_n \mathcal{T}_n
\label{eq:LB}
\end{equation}

\noindent where each channel contributes a unit of conductance multiplied by the transmission probability $\mathcal{T}_n$  of the channel. In the QPC, the conducting channels are based on the set of transverse electronic wave functions, whereas the conducting channels are supported by a set of states located near the edges of the conductor in the IQHE. It is especially interesting to observe once again that the resistance of a ``perfect'' single channel conductor (a QPC with a fully transmitted channel $\mathcal{T}=1$) is not $0$, but is actually equal to the resistance quantum $R_K=h/e^2$. The origin of this residual resistance usually refers to the sum of the two contact resistances $R_c=h/2e^2$, one for each reservoir-lead interface \cite{IBM_1_233_PhilMag_21_863_Landauer,Phys_Mes_Imry_86,Z_Phys_68_217_Landauer,PRB_23_4828_Anderson,PRL_57_1761_Buttiker}. Going back to the  \textit{Gedankenexperiment}, we see that the electronic wave function can only relax in one reservoir lead. Given the context, we examine the following questions: Is it sufficient to claim that the charge relaxation resistance will be equal to the contact resistance? Which kind of transmission dependence is expected for the charge relaxation resistance? To answer these questions, we realized the quantum equivalent of an RC circuit in a 2DEG by associating a quantum point contact (QPC) and a submicrometer quantum dot (QD) (see figure \ref{fig2}). The QD is capacitively coupled to a metallic electrode and can exchange electrons only with an electron reservoir via a single channel of the QPC. At the outset, one might expect that this circuit is the simple association of the QPC resistance $h/(e^2\mathcal{T})$ in series with a geometric capacitance $C$ (see figure \ref{fig2} top). However, the quantum RC circuit is a fully phase-coherent system, where interferences between the QPC and the QD lead to discrepancies with its classical counterpart. Although the RC time can still be written as the product of resistance and capacitance in the low frequency regime $\omega \tau_{RC} \ll 1$, A. Pr\^{e}tre, H. Thomas and M. B\"{u}ttiker predicted in Ref. \cite{Phys_Lett_A180_364_Buttiker,PRL_70_4114_Buttiker,PRB_54_8130_Buttiker} that, for a single spin-polarized quantum channel, the quantum RC circuit would involve a constant charge relaxation resistance $R_q=R_c=h/2e^2$ and a transmission-dependent electrochemical capacitance $C_{\mu}$. More precisely, the capacitance $C_{\mu}$ is the serial combination of the geometric capacitance $C$ and the quantum capacitance $C_q$ \cite{PRL_68_3088_Ashoori,PRL_71_613_Ashoori} resulting from the Pauli exclusion principle in the QD.

The most remarkable result we observed was the universal charge relaxation resistance quantization at half of the resistance quantum $R_K$ at arbitrary transmission \cite{Phys_Lett_A180_364_Buttiker,EPL_37_441_Buttiker,PRB_57_1838_Buttiker,Nat_Phys_1_690_Mora}. While the dc resistance of a single channel conductor is limited by twice the contact resistance, the charge relaxation resistance always equals this contact resistance and corresponds to the dissipation experienced by the quantum RC circuit \cite{PRB_45_3807_Buttiker}. Aside from the case of the RC circuit, the quantum charge relaxation resistance including its generalization in non-equilibrium systems, is an important concept that can be applied to a large number of situations. For example, it is highly relevant to the study of very different problems such as the quantum-limited detection of charge qubits
\cite{PRL_89_200401_Buttiker,PRB_67_165324_Clerk,PRL_102_236801_Nigg},
the study of high-frequency-charge quantum noise
\cite{PhysRep_89_200401_Buttiker,PRL_96_056603_Hekking}, of the dephasing in an electronic quantum interferometer
\cite{PRB_68_161310_Buttiker}, or in the study of electronic
interactions in quantum conductors \cite{PRL_81_1925_Buttiker}. In
molecular electronics, the charge-relaxation resistance is also
relevant to the THz frequency response of systems such as carbon
nanotubes \cite{IEEE_55_Burke}.

In this review, we focus on the experimental realization of the quantum RC circuit and the measurement of the charge relaxation resistance \cite{Science_313_499_Gabelli}. Before describing the experiment, in Section \ref{section2}, we first derive the dc and ac conductances of a two-contact coherent conductor by applying the scattering theory of Landauer-B\"{u}ttiker. The comparison between the two emphasizes the importance of the self-consistent approach to obtaining current conserving expressions for frequency-dependent conductances. Following these general considerations, in Section \ref{section3}, we propose a scattering model for the coherent RC circuit and compare it with the experimental results. Our experiments demonstrate that the series association of a quantum capacitor and a single channel quantum resistor leads to a constant charge relaxation resistance of $h/2e^2$. In Section \ref{section4}, we describe the experimental setup for the measurement of the in-phase and out-of-phase parts of the linear ac conductance of a coherent conductor. Despite its simplicity, the non-interacting and full coherent model presented in this review is remarkably useful in providing an elementary understanding of our experimental results. Moreover, these first experimental investigations have sparked a growing interest in more realistic regimes, both theoretically and experimentally. A brief overview of these works will be given in the conclusion.

\section{Finite frequency conductance in coherent conductors - the RC circuit} \label{section2}

\subsection{General considerations}

In applying the Landauer-B\"{u}ttiker theory, one can describe how the current flows through mesoscopic conductors that are connected to the electron reservoirs \cite{Imry_book,Datta_book,Nazarov_book}. In such conductors (nano-structures at sufficiently low temperature), the coherent length is larger than the typical size of the conductor, such that the wave nature of electrons plays a significant role in the transport. The non-equilibrium steady state currents are thus described by means of the scattering matrix $s(\epsilon)$ that encodes the scattering of non-interacting electronic waves at energy $\epsilon$ in the conductor leads. The electron populations of the incoming states of the leads are imposed by the electron reservoirs and given by the Fermi-Dirac distribution $f$ with an electrochemical potential fixed by the reservoir. As an example, consider a spin polarized single electronic mode transmitted from a left reservoir to a right reservoir (see figure \ref{eq:g_dc}). If the conductor is dc voltage-biased, electrochemical potentials are given by $\mu_L= \epsilon_F-eV$ for the left reservoir, and $\mu_R=\epsilon_F$ for the right one, where $\epsilon_F$  is the Fermi energy and $V$  the dc voltage. The dc conductance $g_{dc} \equiv I/V$ is then given by:

\begin{equation}
g_{dc} = \frac{e^2}{h} \int d\epsilon \, \left(1-s_{LL}^{\star}(\epsilon)s_{LL}(\epsilon)\right) \frac{f(\epsilon)-f(\epsilon+eV)}{eV}
\label{eq:g_dc}
\end{equation}

\noindent $s_{LL}(\epsilon)$ is the amplitude of the probability to be reflected from the left reservoir to itself at
energy $\epsilon$, such that $s_{LL}^{\star}(\epsilon)s_{LL}(\epsilon)$ in equation (\ref{eq:g_dc}) is the reflection probability $\mathcal{R}$. Note
that we have artificially broken the symmetry between the two reservoirs: the current $I=I_{LL}-I_{LR}$ measured in the left reservoir corresponds to the difference between the current coming from the left reservoir $I_{LL} \propto (1-s_{LL}^{\star}(\epsilon)s_{LL}(\epsilon)) \,f(\epsilon -
\mu_L)$ and the current coming from the right one $I_{LR} \propto s_{LR}^{\star}(\epsilon)s_{LR}(\epsilon) \, f(\epsilon - \mu_R)$
where $s_{LR}^{\star}(\epsilon)s_{LR}(\epsilon)$ is the transmission probability $\mathcal{T} =1- \mathcal{R}$. At small excitation ($eV \ll k_BT$) and in the case that the scattering matrix does not depend on the energy scale $k_BT$, we recover the Landauer-B\"{u}ttiker formula $g_{dc}=\frac{e^2}{h} \, \mathcal{T}$ where $h/e^2 \simeq 25.8 \, \mathrm{k} \Omega$ is the quantum of resistance. Altogether, this formalism has proven to be an essential and invaluable tool for theoretically investigating phase-coherent electron transport  \cite{Imry_book,Datta_book,Nazarov_book} and understanding the relevant experiments \cite{PRL_60_848_Wees}.

\begin{figure}[h]
\begin{center}
\includegraphics[width=0.6\linewidth]{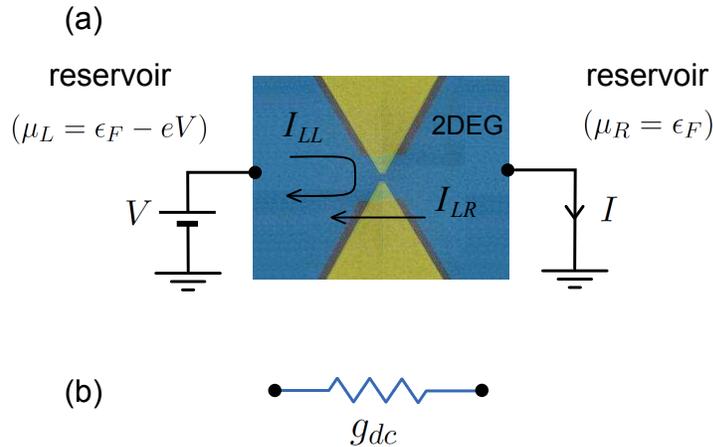}
\end{center}
\caption{(a) Scanning electron microscope view of a quantum point contact (QPC): the 2DEG (blue) is connected to ohmic contacts (not shown) well described by the Landauer-B\"{u}ttiker reservoirs. (b) Schematics of the QPC circuit.} \label{fig3}
\end{figure}

\subsection{Dynamic conductance in the Landauer - B\"{u}ttiker formalism}

Although most of the experiments focused on dc measurement, M. B\"{u}ttiker and his group also addressed the question of the ac conductance in the early 1990s \cite{PRL_70_4114_Buttiker,PRL_71_465_Buttiker}. Indeed, mesoscopic conductors can be driven out of equilibrium by applying oscillating voltages $V_{ac}$ at frequency $\omega$. Firstly, what makes investigating frequency-dependent transport interesting is that one expects the ac conductance to directly probe the intrinsic time scales of the conductor. Secondly, at non zero frequency, current is no longer given by the steady state currents in the coherent conductor and thus, displacement current has to be considered. We will therefore show in the following that the ac conductance gives access to:

 \begin{itemize}
   \item[-] the dwell time $\tau_d$ of the electrons in the mesoscopic capacitance, and
   \item[-] the characteristic charge relaxation time $\tau_{RC}$ that takes into account Coulomb interaction effects in the mesoscopic capacitance.
 \end{itemize}

\begin{figure}[h]
\begin{center}
\includegraphics[width=0.6\linewidth]{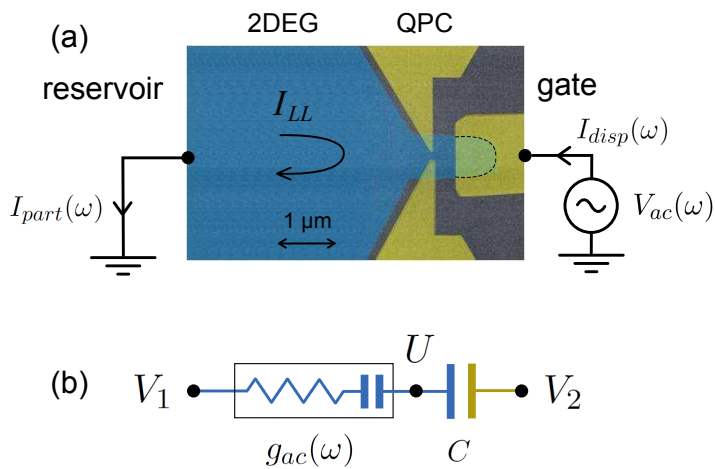}
\end{center}
\caption{(a) Scanning Electronic Microscope view of the sample: the 2DEG (blue) is connected to an ohmic contact (not shown) on the left side and is capacitively coupled to a top gate (yellow) on the right side. The quantum point contact (QPC) (yellow side gates) is used to tune the transmission $\mathcal{T}$  of the electronic wave. (b) Schematics of the mesoscopic RC circuit: the ac conductance $g_{ac}(\omega)$ of the coherent 2DEG is in series with the geometric capacitance $C$.} \label{fig4}
\end{figure}

\noindent Now consider a general mesoscopic RC circuit made of a mesoscopic cavity connected via a single lead to a reservoir and capacitively coupled to a gate (see figure \ref{fig4}). $V_1=0$ is the potential of the reservoir, while $V_2=V_{ac}\cos \omega t$ is the time-dependent potential applied to the gate, and $U$ is the electrostatic potential in the cavity. For simplicity, we assume that the potential $U$ in the cavity is uniform. To determine the ac conductance resulting from the particle current coming from the reservoir, it is necessary to set the potential in the cavity, which is \textit{a priori}  time-dependent. According to gauge invariance, an overall potential shift $-U$ cannot have any effect on the system. We can therefore consider that the reservoir and the gate have oscillating potentials $-U$ and $V_{ac}-U$  while the potential in the cavity is set to zero. The particle current $I_{part}(\omega)$ flowing in the electronic reservoir can then be calculated, thus defining the conductance $g_{ac}(\omega)$  relating to the oscillating potential $-U$ applied to the reservoir. But in ac transport, the particle current alone does not satisfy current conservation. This is particularly clear in the single terminal geometry considered here and depicted on figure \ref{fig4}. To recover current conservation, it is necessary to consider the displacement current $I_{disp}(\omega)$ that flows into the metallic armature of the gate and which equals the particle current flowing into the quantum conductor: $I_{disp}=I_{part}=I(\omega)$. In this respect, We are interested in the
conductance $G(\omega)$ of the whole circuit, connecting the
current $I(\omega)$ to the voltage drop across the whole circuit
$V_{ac}$, and that needs to be distinguished from
$g_{ac}(\omega)$, the conductance of the coherent part of the
circuit. Thus, the calculation of $G(\omega)$ may be performed in two steps:

\vspace{0.3cm}
\noindent (i) Calculating $g_{ac}(\omega)$ for non-interacting electrons; and

\vspace{0.2cm}
\noindent(ii) Self-consistently determining the voltage drop between the cavity and the gate with the help of current conservation. Interactions between the charges in the cavity and the gate are treated in the mean field approximation using the geometric capacitance $C$.
\vspace{0.3cm}

\noindent  Regarding step (i), as far as a single mode conductor with a single reservoir is concerned (see figure \ref{fig4}), the ac conductance changes drastically when compared to the dc conductance. Although there is no dc conductance in this case ($g_{dc}=0$ because $\mathcal{R}=s_{LL}^{\star}(\epsilon)s_{LL}(\epsilon)=1$ in equation (\ref{eq:g_dc})), the ac-conductance $g_{ac}(\omega)=I(\omega)/V_{ac}(\omega)$ is given by:

\begin{equation}
g_{ac}(\omega) = \frac{e^2}{h} \int d\epsilon \, \left(1-s_{LL}^{\star}(\epsilon)s_{LL}(\epsilon+\hbar \omega)\right) \frac{f(\epsilon)-f(\epsilon+\hbar \omega)}{\hbar \omega}
\label{eq:g_ac}
\end{equation}

\noindent To highlight the role of finite frequency and the validity domain of equation (\ref{eq:g_ac}), let us compare it to equation  (\ref{eq:g_dc}). First, we noticed that the energy $eV$, resulting from the dc voltage bias, is replaced here by $\hbar \omega$. Here the bias voltage imposed on the reservoir is $-U$. Due to the linear response of the capacitance $C$, $-U$ is time-dependent and oscillates at the same frequency as $V_2=V_{ac} \cos \omega t$. Thus, an ac bias voltage $-U \cos \omega t$  is imposed on the reservoir and the electron wave functions acquire an extra phase factor $\sum_n J_n(eU/\hbar \omega) \exp(i n \omega t)$, where $J_n$ is the ordinary Bessel function \cite{PR_129_647_Tien}. In the low voltage limit, $eU/\hbar \omega \ll 1$, this phase factor reduces to $1+(eU/2 \hbar \omega) \,\exp(i\omega t)-(eU/2 \hbar \omega)\, \exp(-i\omega t)$ at the first order in $eU/\hbar \omega$. Thus, the current $I_{LL}(\omega)$ oscillating at frequency $\omega$ arises from the interference of processes, where electrons absorb or emit one photon at energy $\hbar \omega$; each absorption/emission process is weighted by an amplitude $\pm
eU/2 \hbar \omega$. As a result, $I_{LL}(\omega)$ exhibits the Fermi- Dirac factors $(f(\epsilon)-f(\epsilon+\hbar \omega))/\hbar
\omega$. Furthermore, we noticed that the ac current, unlike the dc one, is clearly related to the intrinsic dynamics of the conductor \textit{via} the term $s_{LL}^{\star}(\epsilon)s_{LL}(\epsilon+\hbar \omega)$ \footnote{Equation (\ref{eq:g_ac}) does not consider electron-electron interactions in the reservoir. It is valid insofar as $\omega$ is smaller than frequencies associated with electrodynamics, \textit{i.e.} the plasma frequency of the 2DEG. In our experiment, $\omega/ 2\pi \sim 1 \, \mathrm{GHz}$ and $\omega_p/2\pi \sim 1 \, \mathrm{THz}$  at $B=1 \, \mathrm{T}$, leading to  $\omega/\omega_p \sim 10^{-3}$\cite{PRB_84_041305_Baskin}.}.

Regarding step (ii), in the process of self-consistently determining $U$,  the displacement current $I_{disp}$ is the time-derivative of the charge in the cavity, which has to be identified to particle current $I_{part}=I_{disp}=I(\omega)$ in order to recover the current conservation\footnote{The complex representation of voltage drop across a capacitance is conventionally $V=I/(iC \omega)$, where angular frequency is positive.}:

\begin{equation}
g_{ac}(\omega) \, U=  iC\omega \, \left( V_{ac} -U\right)
\label{eq:Ipart}
\end{equation}

\noindent Finally,  $G(\omega)=I(\omega)/V_{ac}(\omega)$ is given by:

\begin{equation}
G(\omega)=\frac{1}{1/g_{ac}(\omega)+1/(iC\omega)}
\label{eq:Ggen}
\end{equation}

\noindent The total conductance is equivalent to the series addition of the geometric capacitance $C$ and the impedance $1/g_{ac}$ of the coherent conductor. Although we considered a single mode conductor in this case, both equations (\ref{eq:g_ac}) and (\ref{eq:Ggen}) can be easily extended to the multi-mode case by considering $g_{ac}$ as the parallel addition of channel conductances \cite{PRB_54_8130_Buttiker}.

\subsection{Charge relaxation resistance}

At finite frequency, the conductance is a complex number, made up of a real part (the conductance), and an imaginary part (the susceptance). These two quantities are a combination of dissipative elements (\textit{e.g.} resistors) and reactive elements (\textit{e.g.} capacitors). To precisely define the nature of the ac conductance $G(\omega)$, it is necessary to expand it to second order in frequency. The first order term in the current response of the whole circuit gives access to the electrochemical capacitance $C_{\mu}$ of the circuit, whereas the second order term yields both the relaxation time $\tau_{RC}=R_q \times C_{\mu}$  and the charge relaxation resistance  $R_q$:

\begin{equation}
G(\omega)= i \omega C_{\mu} \left(1-i \omega \tau_{RC} + \mathcal{O}\left((\omega\tau_{RC})^2\right) \right) \label{eq:G}
\end{equation}

\noindent  We now turn to expressing $C_{\mu}$ and $R_q$  with respect to the scattering matrix $s(\epsilon)$ related to the coherent cavity coupled to the reservoir. For the sake of simplicity, let us consider the low temperature limit, where the Fermi-Dirac factors in equation (\ref{eq:g_ac}) is simplified to a delta function. Since we deal with a scattering problem that exclusively involves reflections, $s(\epsilon)$ solely relates to the phase $\phi_n$ that an electron accumulates in the $n$th channel mode of transmission: $s_n(\epsilon)=e^{i \phi_n(\epsilon)}$. Thus, applying the derivation into the second order in frequency of equation (\ref{eq:g_ac}), and considering addition of parallel $n$ channels, we identify:

\begin{eqnarray}
C_{\mu}=\frac{CC_q}{C+C_q}  \label{eq:Cmu}\\
C_q=e^2N(\epsilon_F) \label{eq:C_q}\\
R_q=\frac{h}{2e^2} \, \frac{\sum_n \tau_n ^2}{\left(\sum_n \tau_n \right)^2}
\label{eq:C_qR_q}
\end{eqnarray}

\noindent where $\tau_n= \hbar \frac{d \phi_n}{d\epsilon}(\epsilon_F)$  is the time an electron in the $n$th channel spends in the cavity, $\tau_d = \sum_n \tau_n$ is the dwell time of the electrons in the mesoscopic capacitance and  $N(\epsilon_F)=\tau_d/h$ is the local density of states in the  mesoscopic cavity. The above expressions are valid when the third order in frequency can be disregarded, \textit{i.e.} for $\omega
\tau_d \ll 1$ (the development of the conductance up to the third order corresponds to the addition of an inductive contribution in series with the resistance and capacitance \cite{PRB_75_155336_Wang}). According to equations (\ref{eq:Cmu}) and (\ref{eq:C_q}), the capacitance of the mesoscopic structure is found to be a series combination of the geometric capacitance $C$ and the quantum one $C_q$ \cite{PRL_68_3088_Ashoori}. Moreover, we see in equation (\ref{eq:C_qR_q}) that the charge relaxation resistance is quantized in a peculiar way and does not depend directly on the transmission. In the case of a single spin-polarized channel,  it reduces to half of the quantum of
resistance and is not transmission-dependent:

\begin{equation}
R_q= \frac{h}{2e^2}
\label{eq:R_q}
\end{equation}

\section{Charge relaxation in the coherent RC circuit}\label{section3}

\subsection{Scattering theory and density of states of a mesoscopic capacitor}

\begin{figure}[!htph]
\centering\includegraphics[scale=0.8]{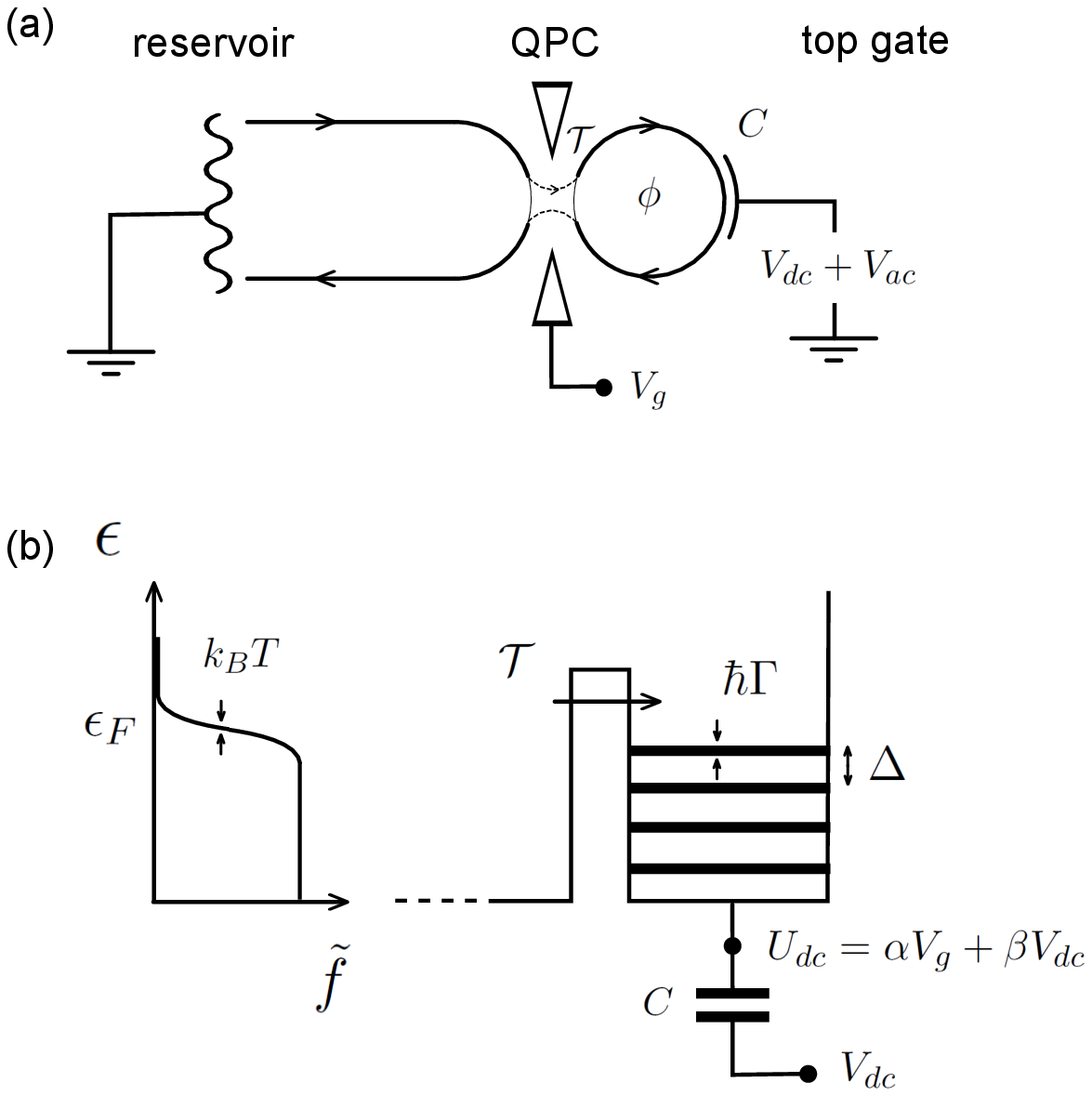}
\caption{(a) One-dimensional model of the mesoscopic RC circuit. The quantum dot (QD) is tunnel coupled (transmission $\mathcal{T}$) to a two-dimensional electron gas through a quantum point contact (QPC). An electronic wave acquires the phase $\phi$ in a single roundtrip in the QD. The transmission $\mathcal{T}$ is voltage-controlled by a gate voltage $V_g$, while dc and ac voltages ($V_{dc}$ and $V_{ac}$) can be applied to the top gate. (b) Energy levels in the QD are quantized at small transmission $\mathcal{T}$. The width $\hbar\Gamma$  of the energy level is transmission-dependent, whereas the level spacing $\Delta$ depends on the geometry of the QD.}
\label{fig5}
\end{figure}

As discussed in Section \ref{section2}, the quantum RC circuit depicted in figure \ref{fig4}  consists of a submicron-sized electronic cavity (or quantum dot) tunnel coupled to a two-dimensional electron gas through a quantum point contact (QPC), whose transmission $\mathcal{T}$  is controlled by the gate voltage $V_\mathrm{g}$ and capacitively coupled to a macroscopic electrode deposited on top of the 2DEG. In this quantum version of the RC circuit, the dot and electrode define the two plates of a capacitor while the quantum point contact plays the role of the resistor. A large perpendicular magnetic field is applied to the sample in order to reach the Integer Quantum Hall regime, where the electrons propagate ballistically along the chiral edge channels. We consider the situation where a single edge channel is coupled to the dot, such that electronic transport can be described by the propagation of spinless electronic waves in a one-dimensional conductor. Thus, electrons in the incoming edge channel can tunnel into the quantum dot with the amplitude, $\sqrt{\mathcal{T}}=\sqrt{1-r^2}$, perform several roundtrips inside the cavity, each taking the finite time, $\tau_0 =l/v_d$, before finally tunneling back out into the outgoing edge state (see figure \ref{fig5}(a)).  For the sake of convenience, the reflection amplitude $r$ in these expressions has been assumed to be real and energy-independent, while  $l$ and $v_d$ represent the circumference of the quantum dot and the drift velocity, respectively. For a micron size cavity, $\tau_0$ typically equals a few tens of picoseconds.

As shown in Section \ref{section2}, the dynamic properties of the circuit are encoded in the scattering matrix $s(\epsilon)$ describing the scattering of an electronic wave at energy $\epsilon$  by the quantum dot. More precisely, using equation (\ref{eq:g_ac}), the ac conductance of the coherent part of the circuit can be fully characterized knowing its scattering properties $s(\epsilon)$. In the geometry considered here, the quantum dot acts as the electronic analog to a Fabry- P\'erot cavity. An electronic wave of energy $\epsilon$ acquires the phase $\phi(\epsilon)=(\epsilon -eU_{dc}) \tau_0/\hbar$  in a single round trip in the cavity, with $U_{dc}$ being the static potential of the dot. The scattering matrix $s(\epsilon)$ can then be easily computed as the sum of the amplitudes for all the processes required to generate an arbitrary number of round trips inside the cavity:

\begin{eqnarray}
s(\epsilon) & = &  r - \mathcal{T} e^{i \phi(\epsilon) } \sum_{q =0
}^{\infty} r^q e^{i q \phi(\epsilon)} \\
s(\epsilon) & = & \frac{r -e^{i \phi(\epsilon)}}{1 - r^{i
\phi(\epsilon)}} = e^{i \Theta(\epsilon)}
\label{eq:s}
\end{eqnarray}

\noindent As expected, the scattering matrix is a pure phase, as the electrons entering the cavity
leave it with unit probability. The density of states of the cavity can be immediately
deduced, $N(\epsilon)=\frac{1}{2i \pi} s^{*} \frac{ds}{d \epsilon} = \frac{1}{2 \pi }\frac{d \Theta}{d\epsilon}$:

\begin{eqnarray}
N(\epsilon) & = & \frac{\tau_0}{h} \frac{1 -r^2}{1 -2r \cos\left(\frac{2\pi}{h}
(\epsilon-eU_{dc}) \tau_0 \right) + r^2}
\label{eq:N}
\end{eqnarray}

\noindent As expected, at unit transmission ($r=0$), the density of states in the dot is uniform and related to the time spent in the cavity in a single round trip, $N(\epsilon) = \tau_0/h$. When the transmission is decreased, the density of states exhibits a periodic structure with period $\Delta = h/\tau_0$ that reflects the resonant tunneling inside the cavity (where $\Delta/k_B$ typically equals a few Kelvins). In the limit of small transmissions ($\mathcal{T} \ll 1$, $r\approx 1$), the density of states is a sum of lorentzian peaks of width $\hbar \Gamma$, $\Gamma = \mathcal{T} /\tau_0$ (see figure \ref{fig5}(b)):

\begin{eqnarray}
N(\epsilon) & \approx &\frac{2}{\pi \hbar \Gamma} \,\sum_{n}
\frac{1}{1+ \left(\frac{\epsilon-eU_{dc} - n \Delta}{\hbar \Gamma/2}\right)^2}
\end{eqnarray}

\noindent In this limit where the dot is weakly coupled to the one-dimensional edge channel, these peaks can be viewed as the discrete spectrum of the dot energy levels. By increasing the transmission, the width of the levels increases up to the point where the peaks fully overlap, thereby forming a continuous spectrum. Note that by tuning the dot static potential $U_{dc}$, the dot spectrum can be shifted with respect to the Fermi energy of the edge channel $\epsilon_F$ (see figure \ref{fig5}(b)). As shown by equation (\ref{eq:G}) in Section \ref{section2}, the circuit can be represented at low frequency by the serial addition of the geometric capacitance  $C$, a quantum capacitance $C_q$ (for a total capacitance $C_{\mu}$) and a charge relaxation resistance $R_q$. The quantum capacitance and charge relaxation resistance are directly related to the scattering properties of the conductor by:

\begin{eqnarray}
C_q & = &  e^{2}\int d\epsilon N(\epsilon) \left(-\frac{df}{d\epsilon}\right) \label{EqCqT}\\
R_q & = & \frac{h}{2e^2} \frac{\int d\epsilon N(\epsilon)^2
\left(-\frac{df}{d\epsilon}\right)}{\left(\int d\epsilon N(\epsilon)
\left(-\frac{df}{d\epsilon}\right)\right)^2} \label{EqRqT}
\end{eqnarray}

\noindent When the density of states varies smoothly on the scale of the electronic temperature $k_BT$, the effects of temperature can be ignored. We then recover the zero temperature expressions of equations (\ref{eq:C_q}) and (\ref{eq:R_q}) of the quantum capacitance $C_q=e^2 N(\epsilon_F)$ and the charge relaxation resistance $R_q=h/(2e^2)$. The expression of the capacitance can be easily understood. When the dot potential is varied by $dU$, due to the finite density of states, the number of charges that can enter the dot is $dQ = e^2 N(\epsilon_f) dU$. The quantum capacitance thus provides a direct spectroscopy of the discrete energy levels of the dot. As such, it is sensitive to all the dot parameters, the dot potential $U_{dc}$, and in particular the transmission $\mathcal{T}$. The most striking effect of phase coherence appears on the charge relaxation resistance, which is quantized to the universal value of $R_q=h/(2e^2)$ and does not depend on the dot transmission $\mathcal{T}$. The relaxation resistance that appears in the dynamics of charge transfer strongly differs from the dc resistance $R_{dc}$ of a conductor transmitting with transmission $\mathcal{T}$ a single spinless channel between two electronic reservoirs. In the latter case, the resistance depends on the transmission through the Landauer formula $R_{dc}= h/(\mathcal{T}e^2)$. Moreover, at unit transmission, it is quantized to the value $h/e^2$, which is twice the value of the charge relaxation resistance. This factor of two can be explained by the presence of two electronic reservoirs instead of only one for the charge relaxation resistance.

Using equation (\ref{eq:G}), the quantum capacitance $C_q= e^2
N(\epsilon_F)$ and the charge relaxation resistance $R_q=h/(2e^2)$ can be directly probed by the measurement of the imaginary and real parts of the conductance $G$. The low frequency behavior described in equation (\ref{eq:G}) is valid for $\tau_{RC} = R_{q} C_{\mu } \omega \ll 1$, which is satisfied when $\omega \tau_d \ll 1$. However, to get an accurate measurement of the charge relaxation resistance, the angular frequency $\omega$ has to be selected, such that the real part of the conductance is not vanishingly small. For $\tau_d$ of the order of a few tens of picoseconds, these conditions are satisfied for $\mathrm{GHz}$ frequencies.

\subsection{Experimental determination of the conductance - Impedance of a coherent RC circuit}

\begin{figure}[!htph]
\centering\includegraphics[scale=0.6]{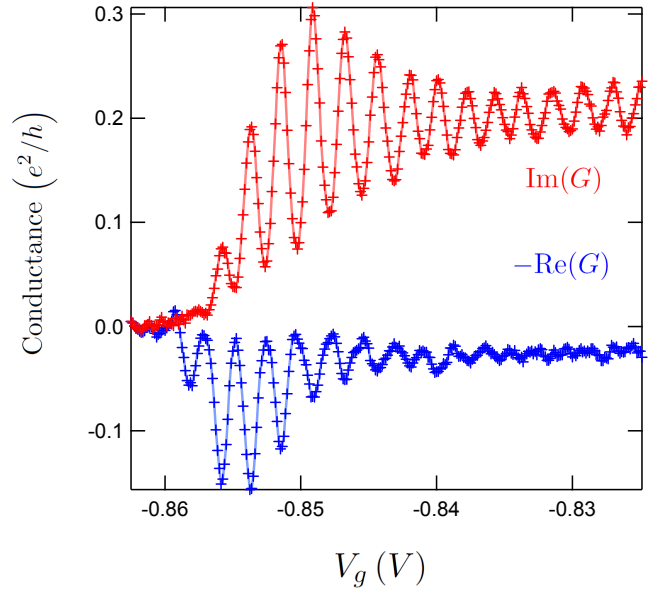}
\caption{Real and imaginary parts of the conductance $G$ in sample S3 measured at a frequency of $\omega/2\pi=1.24 \, \mathrm{GHz}$.} \label{fig6}
\end{figure}

Figure \ref{fig6} presents both the imaginary and real parts of the conductance of a first sample, labeled S3, as a function of the gate voltage $V_\mathrm{g}$, which is applied to the quantum point contact. The measurements are performed at the frequency of $\omega/2\pi=1.2\ Ghz$. $V_\mathrm{g}$  has two effects on the dot parameters. Firstly and most apparently, it controls the transmission $\mathcal{T}(V_\mathrm{g})$ from the cavity to the one-dimensional edge channel. By tuning the gate voltage to negative values (starting from the right side of figure \ref{fig6}), the full range of transmissions can be accessed from a perfectly open cavity at $\mathcal{T}=1$ to a fully closed cavity $\mathcal{T}=0$. The second effect is to linearly change the static potential of the dot $U_{dc}=\alpha V_\mathrm{g}$ by a capacitive coupling from the QPC gate to the dot (the dot is also capacitively coupled to the top gate, such that the general expression of the static dot potential is $U_{dc}=\alpha V_\mathrm{g} + \beta V_{dc}$, where $V_{dc}$ is the static potential applied to the gate). In the range of $V_g \geq -0.845 \, \mathrm{V}$, which corresponds to high transmissions, the imaginary part of the conductance is much greater than the real part, in accordance with the low frequency description of the circuit: $R_q C_{\mu}\omega \ll 1$. The imaginary part presents pronounced oscillations with the gate voltage, which are less visible on the real part. These oscillations correspond to the modulation of the dot density of states when the dot potential $U_{dc}$ is varied, see equation (\ref{eq:N}). When a dot energy level is resonant with the Fermi energy, the capacitance (and hence the imaginary part of the conductance) exhibits maximum oscillations. However, when a dot energy level is out of resonance with the Fermi energy, the capacitance exhibits minimum oscillations.

When $V_\mathrm{g}$ is decreased, $V_{g} \leq -0.845 \, \mathrm{V}$, the real part of the conductance increases and becomes comparable with the imaginary part (for $V_{g} \approx -0.855 \, \mathrm{V}$). By decreasing the gate voltage further, the conductance eventually vanishes, as one might expect in the pinched situation where the transmission is close to zero. In this limit, $V_{g} \leq -0.855 \, \mathrm{V}$, the signal is mainly carried by the real part, $\mathrm{Re}(G) \gg \mathrm{Im}(G)$.

\begin{figure}[!htph]
\centering\includegraphics[scale=0.6]{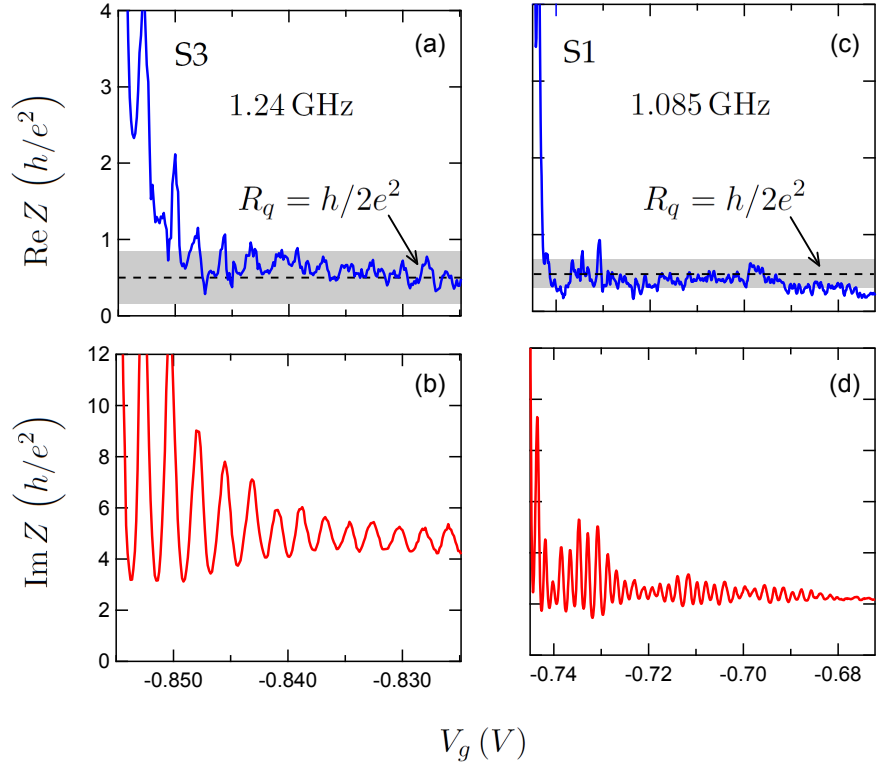}
\caption{Real and imaginary parts of the impedance $Z$ in samples S3 (left panel) and S1 (right panel). Measurements in sample S3 were performed at a frequency of $\omega/2\pi=1.24 \, \mathrm{GHz}$, while measurements in sample S1 were performed at $\omega/2\pi=1.085 \, \mathrm{GHz}$.}
 \label{fig7}
\end{figure}

Keeping our focus on the regime of high transmissions in figure \ref{fig6}, $V_g \geq -0.845 \, \mathrm{V}$ and $\mathrm{Im}(G) \gg \mathrm{Re}(G)$, we  also consider on figure  \ref{fig7} the real and imaginary parts of the impedance $Z=1/G$ extracted from our conductance measurements. The data presented on the left panel are extracted from the measurements of figure \ref{fig6}, while the right panel data have been obtained from another sample labeled S1. In the lumped element description of an RC circuit, the imaginary part is, up to the pulsation, the inverse capacitance, $\mathrm{Im}(Z) = 1/(C_{\mu} \omega)$ while the real part provides a direct measurement of the charge relaxation resistance $\mathrm{Re}(Z) = R_q$. As previously discussed, the oscillations of the capacitance are related to the oscillations of the dot spectrum with respect to the Fermi energy, when $V_g$ is varied. When the transmission $\mathcal{T}$ decreases, the oscillations become more pronounced, as the width of single energy levels $\hbar
\Gamma$ decreases. However, the behavior of the real part of the impedance is completely different. For both samples, the resistance is flat for a wide range of $V_\mathrm{g}$ and its value is given (within error bars) by the expected $R_{q} = \frac{h}{2 e^2}$. In this regime, the resistance is constant and independent of both the dot transmission  $\mathcal{T}$ and static potential $U_{dc}$. These results provide an evidence of the quantization of the charge relaxation resistance of a single mode conductor. However, at some point ($V_g \leq -0.85 \, \mathrm{V}$  for sample S3, $V_g \leq -0.74 \, \mathrm{V}$ for sample S1), the resistance deviates from its universal value and starts increasing rapidly. This increased resistance corresponds to the increase of the real part of the conductance on figure \ref{fig6}, which becomes comparable with the imaginary part when $R_{q} C_{\mu} \omega \approx 1$, and finally dominates for $R_{q} C_{\mu} \omega \gg 1$, when the conductance reaches to zero. This sudden increased resistance, which becomes transmission-dependent in this regime, cannot be explained by equation (\ref{eq:R_q}), since this equation describes the zero temperature behavior of the circuit. To explain this dependence, temperature effects have to be considered in a complete modeling of the circuit.

\subsection{Modeling of the quantum RC circuit at finite temperature}

From equation (\ref{EqCqT}), the quantum capacitance at finite temperature can be written as $C_q=e^2 \tilde{N}(\epsilon_F)$, where  $\tilde{N}(\epsilon_F)$ is an effective density of states resulting from the convolution between the zero temperature density of states $N(\epsilon)$ and the derivative of the Fermi distribution $(-df/d\epsilon)$. At fixed transmission $\mathcal{T}$, the density of states varies on the typical scale $h \Gamma = \mathcal{T} \Delta$  that has to be compared with $k_BT$ ($T \approx 100 \, \mathrm{mK}$). In large transmissions, $\mathcal{T} \approx 1$, $h \Gamma \approx \Delta \gg k_B T$  and the zero temperature description of equation (\ref{eq:C_q}) apply. When $\mathcal{T}$ is decreased however, deviations from the zero temperature expression are expected when $\hbar \Gamma \approx k_BT$. An analytical expression for the resistance and capacitance can be obtained in the limit of $\hbar \Gamma \ll k_B T$:

\begin{eqnarray}
C_q & = &  \frac{e^2}{4 k_B T \mathrm{ch}^2(\frac{\epsilon_n-\epsilon_F}{2 k_BT})} \label{EqCqbraod} \\
R_q & =& \frac{h}{\mathcal{T}e^2} \frac{4 k_B T}{ \Delta}
\mathrm{ch}^2(\frac{\epsilon_n-\epsilon_F}{2 k_BT})\label{EqRqbraod}
\end{eqnarray}

\noindent This limit corresponds to the sequential tunneling regime where the QD is weakly coupled to the reservoir. In this regime, the tunneling process permits many oscillations in the well, \textit{i.e.} a long dwell time, but is dominated by thermal broadening. Then, the charge relaxation resistance is no longer independent of the transmission. Rather, it diverges as $1/\mathcal{T}$ when $\mathcal{T}$ is decreased. Thus, the resistance on figure \ref{fig7} suddenly increases when the gate voltage $V_g$ is decreased. When the resistance increases, such that $R_q C_{\mu} \omega \approx 1$, the low frequency expansion of the conductance (equation (\ref{eq:G})) becomes invalid. As a result, it has to be replaced by the general expressions relating $G(\omega)$ to $g_{ac}(\omega)$ (equation (\ref{eq:Ggen})) and $g_{ac}(\omega)$ to the scattering matrix $s(\epsilon)$ (equation (\ref{eq:g_ac})). Note that, by using our experimental parameters, the general expression of the ac conductance $G(\omega)$ still conforms (up to our experimental resolution) with the conductance of the RC circuit, $G = -iC_{\mu} \omega / (1 + i R_q C_{\mu} \omega)$, whose capacitance $C_{\mu}$ and $R_q$  are still given by their expression deduced from the low frequency behavior in equations (\ref{EqCqT}) and (\ref{EqRqT}). For a quantitative analysis, we rely on the exact expressions of $G(\omega)$, whereby the RC circuit picture has proven extremely useful in providing a qualitative understanding of our conductance trace shown on figure \ref{fig6}. In large transmissions, the zero temperature description remains. When $\mathcal{T}$  is decreased, such that $\hbar \Gamma \approx k_B T$, the real part of the conductance increases, while the imaginary part starts to decrease and the resistance increases. For $R_q C_{\mu} \omega \approx 1$, the real part of the conductance reaches a maximum, whereby as a result, $\mathrm{Re}(G) \approx \mathrm{Im}(G)$, which is expected for an RC circuit. Finally, for the lower values of $\mathcal{T}(V_g)$, the conductance eventually goes to zero and the signal is essentially carried by the real part of the conductance, $\mathrm{Re}(G) \gg \mathrm{Im}(G)$ ($R_q C_{\mu} \omega \gg 1$).

To support this qualitative analysis, a more quantitative description can be performed that relies on equations (\ref{eq:Ggen}),  (\ref{eq:g_ac}) and (\ref{eq:s}) as well as on modeling the effect of $V_g$ on both the transmission $\mathcal{T}(V_g)$ and the static dot potential $U_{dc}(V_g)$. The gate voltage dependence of the transmission depends typically on two parameters: the width $\delta V$, on which the transmission goes from $0$ to $1$, and the gate voltage $V_0$, for which the transmission equals $0.5$. The exact dependence $\mathcal{T}(V_g)$ is taken as the two-parameter Fermi distribution:

 \begin{equation}
 \mathcal{T}(V_g) = \frac{1}{1+ e^{-\frac{V_g-V_0}{\delta V}}}
 \end{equation}

\begin{figure}[!htph]
\centering\includegraphics[scale=0.8]{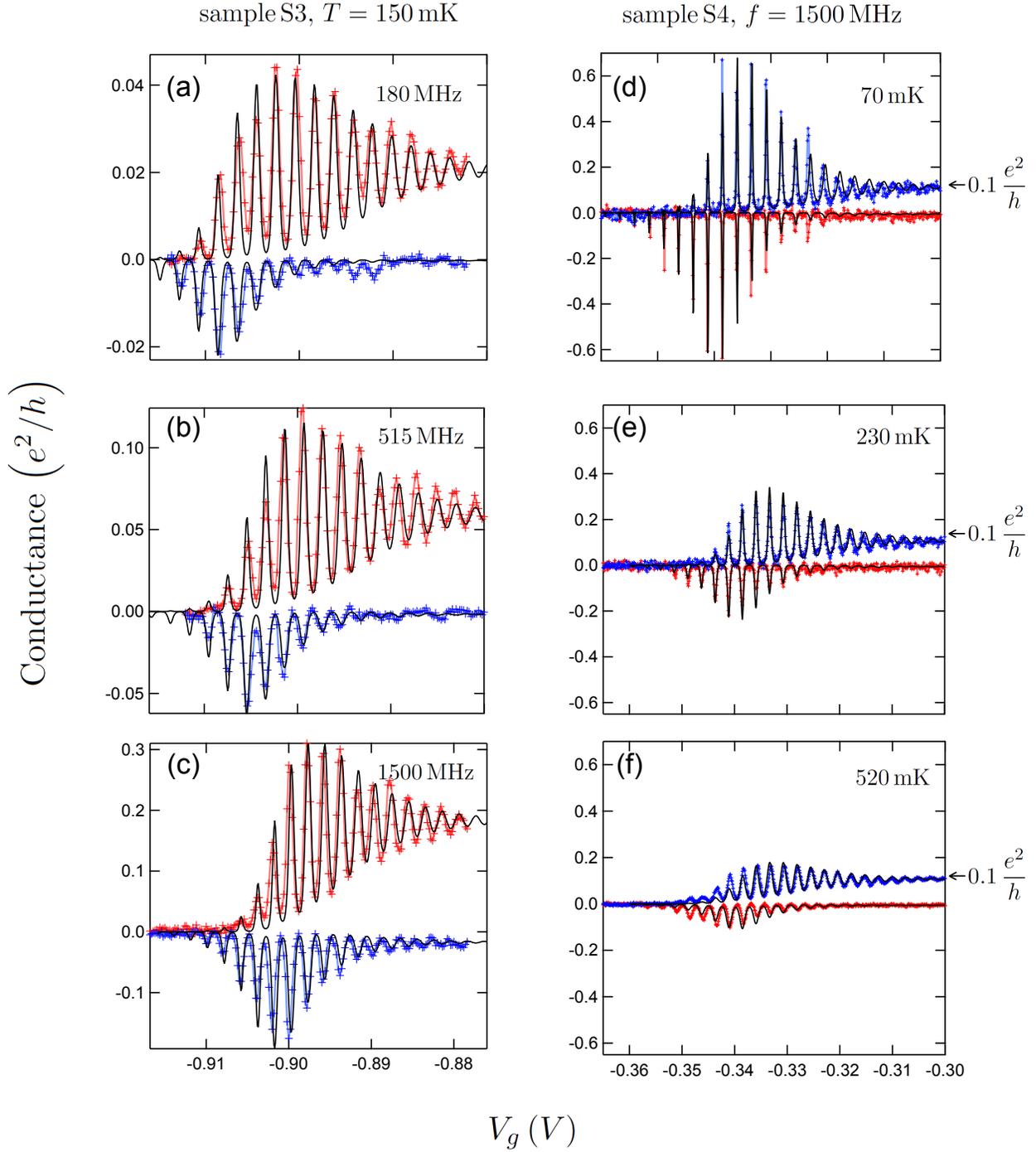}
\caption{Left panel: comparison between measurements (dots) and modeling (traces) of the conductance in sample S3$^{\star}$ at various frequencies of $\omega/2\pi=1.5 \, \mathrm{GHz}$, $515$ and $180 \, \mathrm{MHz}$. Parameters of the model are $T=150 \, \mathrm{mK}$, $C_{\mu} = 0.75 \, \mathrm{fF}$ (measured independently), $C=3.5 \, \mathrm{fF}$, $V_{0}=-896 \, \mathrm{mV}$, $\delta V = 2.9 \, \mathrm{mV}$, and $e\alpha /k_B = 1.2\,
\mathrm{K}.\mathrm{mV}^{-1}$. Right panel: comparison between measurements (dots) and modeling (traces) of the conductance in sample S4 at various temperatures $T=70$, $250$ and $520 \, \mathrm{mK}$ and at frequency $\omega/2\pi=1.5 \, \mathrm{GHz}$. Parameters of the model are $C_{\mu} = 0.44 \, \mathrm{fF}$ (measured independently), $V_{0}=-329.8 \, \mathrm{mV}$, $\delta V = 4.4 \, \mathrm{mV}$, and $e\alpha/k_B = 1.65\,
\mathrm{K}.\mathrm{mV}^{-1}$.} \label{fig8}\end{figure}

\noindent The gate voltage $V_g$ also leads to a shift of the dot potential $U_{dc}$, which we assume to be linear with the gate voltage $U_{dc} = \alpha V_g$ as expected for a capacitive coupling. Using $\mathcal{T}(V_g)$ and $U_{dc}(V_g)$  in equation (\ref{eq:s}), our results can be quantitatively compared with our experimental data with four adjustable parameters, $\delta V$, $V_0$, $\alpha$ and the geometrical capacitance $C$ ($\Delta$ and $T$  being calibrated independently; see Section \ref{section4}). Comparisons can be seen in figure \ref{fig8} , which presents the conductance of another sample labeled S3$^{\star}$ measured at three different frequencies: $\omega/2\pi=1.5 \, \mathrm{GHz}$, $515 \, \mathrm{MHz}$ and $180 \, \mathrm{MHz}$ (left panel). Although the global shape of the conductance traces is not affected when the frequency is decreased, some differences are still noticeable. The maximum of the real part that corresponds to $R_q C_{\mu} \omega \approx 1$ is, as expected, shifted to lower values of $V_g$ that correspond to lower values of $\mathcal{T}$. One also notices that in the high transmission regimes, the real part of the conductance becomes hardly measurable at the lowest frequency, thereby emphasizing the need of GHz frequencies to measure the quantization of charge relaxation resistance, $R_q =h/2e^2$. As can also be seen in figure \ref{fig8}, the data model agreement is excellent, as it uses a single set of parameters for all frequencies. When the frequency is decreased, the shift of the signal to decreased values of the gate voltage as well as the amplitude and positions of the signal oscillations are well captured.

\begin{figure}[h]
\begin{center}
\includegraphics[width=0.6\linewidth]{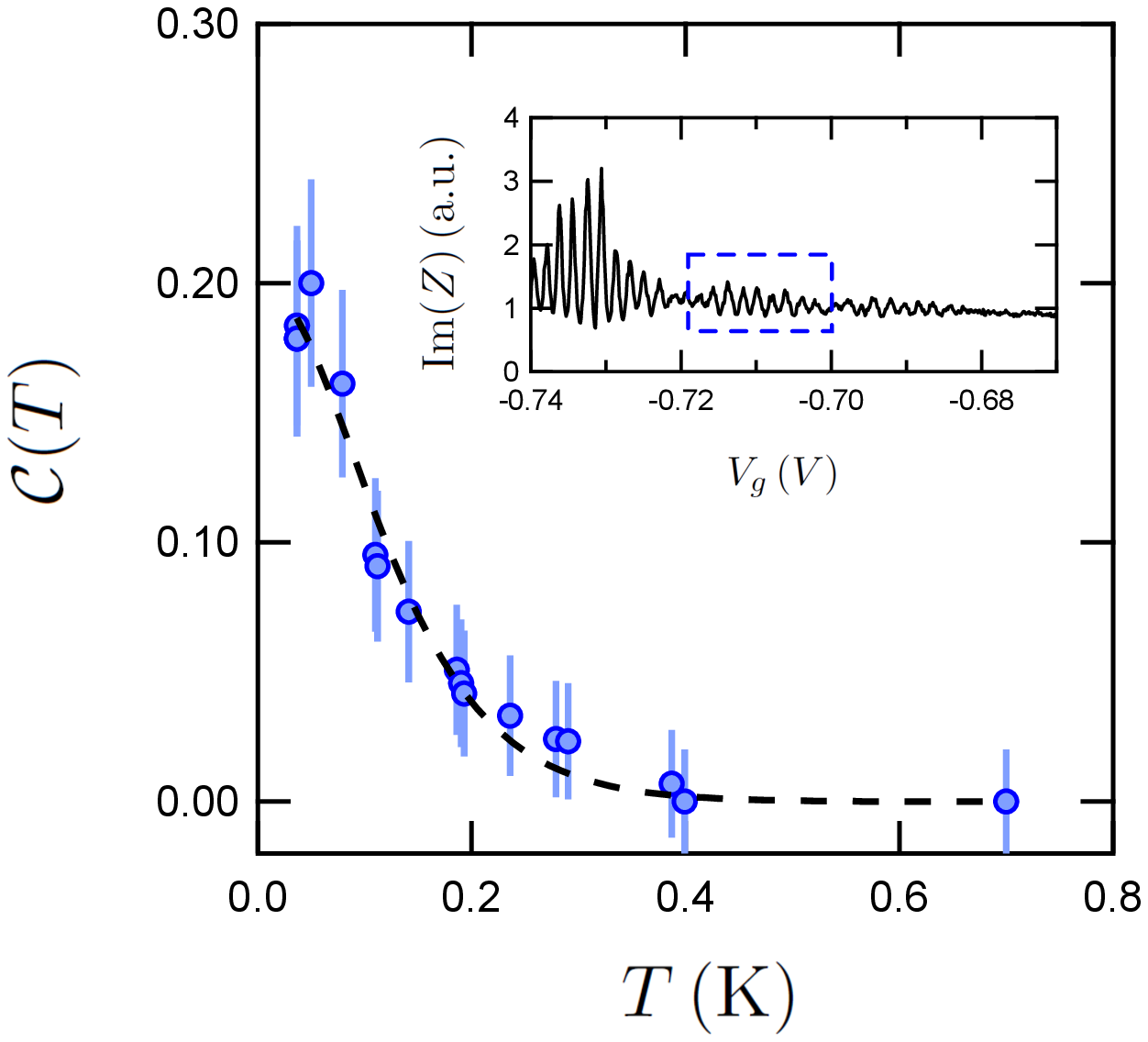}
\end{center}
\caption{Temperature dependence of the amplitude of the capacitance oscillation in sample S1 at $B=1.3 \, \mathrm{T}$ and $\omega/2\pi=1.085 \, \mathrm{GHz}$. The dashed line corresponds to the curve fitting with equation $\mathcal{C}(T)=A \, \left(T/T^{\star}\right)/\sinh
\left(T/T^{\star}\right)$. Inset: normalized imaginary part of the impedance $\mathrm{Im}(Z)$  for $-0.74 \leq V_g \leq -0.68 \, \mathrm{V}$. The dashed rectangle defines the capacitance oscillations, whose temperature dependence is plotted on the main figure.}
\label{fig9}
\end{figure}

The right panel presents the evolution of the conductance with the temperature at fixed frequency (measurements performed on another sample labeled S4) for three different temperatures: $T=70$, $230$ and $520 \, \mathrm{mK}$. At the lowest temperature, the
oscillations of the capacitance are extremely sharp. These
oscillations are strongly affected by the temperature. While the
capacitance at transmission $\mathcal{T}=1$ is not affected, the
maxima of the capacitance are strongly reduced and the peaks width
increases, conformably to expectations. This effect, which directly reflects the dependence of the effective density of states $\tilde{N}(\epsilon)$ on the applied temperature, is perfectly captured by our model for all three temperatures; except for the low transmission part of the $520\, \mathrm{mK}$ trace. This small disagreement might result from a small dependence of transmission on energy that starts affecting our measurements at high temperatures. An exhaustive study of the temperature dependence of the capacitance oscillation in large transmissions was conducted in sample S1 (see figure \ref{fig9}), where experimental points fall again on the theoretical curve deduced from the scattering model:

 \begin{equation}
 \mathcal{C}\equiv\frac{\mathrm{Im} \left(Z_{max}\right)-\mathrm{Im} \left(Z_{min}\right)}{\mathrm{Im} \left(Z_{max}\right)+\mathrm{Im} \left(Z_{min}\right)}=\frac{2r \Delta C_{\mu}}{e} \, \frac{T/T^{\star}}{\sinh T/T^{\star}}
 \end{equation}

\noindent with $k_B T^{\star}= \Delta/(2\pi^2)$. This measurement gives an alternative approach for calibrating the level spacing $\Delta=18 \pm 3 \, \mathrm{GHz}$ for this sample.

This excellent agreement with theoretical predictions, when all parameters are varied (frequency, temperature, dot potential, dot transmission), show that the quantum RC circuit is an optimal model system for studying the dynamic properties of a quantum conductor, whereby these properties can be quantitatively understood.

\subsection{Coulomb interactions and charge relaxation}\label{Coulomb}

So far, Coulomb interactions have been treated in a very simple manner. Their effects have been disregarded in the calculation of the particle current circulating in the conductor, \textit{i.e.} the calculation of $g_{ac}(\omega)$, which determines the conductance of the coherent part of the circuit. Coulomb interactions are only then considered when introducing the geometric capacitance $C$ of the mesoscopic capacitor, which connects the current to the voltage drop between the gate and the mesoscopic cavity: $I(\omega) = i C \omega (V_{ac}-U)$. As a consequence, the quantum capacitance $C_q$, related to $g_{ac}(\omega)$, exhibits periodic peaks that are only related to the single particle level spacing of the dot $\Delta$. However, this description still disregards an important effect of electronic transport in small structures, such as quantum dots, which is also referred to as the ``Coulomb blockade''. Due to the Coulomb interaction, in order to add one electron inside a quantum dot, one has to pay an energetic cost, which is the sum of the orbital level spacing $\Delta$ and the charging energy $e^2/C$ in order to place a charge $e$ on a capacitor plate  $C$ \cite{PRB_51_1743_Matveev,PRB_57_9608_Glazman}. This total energetic cost, which defines a renormalized level spacing $\Delta^{\star}$, is precisely related to the electrochemical capacitance, $\Delta^{\star}= e^2/C + \Delta = e^2/C_{\mu}$. Clearly, a proper account of Coulomb blockade effects in the calculation of $g_{ac}(\omega)$ would modify the dot density of states (and hence the quantum capacitance $C_q$) by introducing the new scale $\Delta^{\star}$. Additionally, one might also wonder if this proper account of Coulomb interactions might break the quantization of the charge relaxation resistance $R_q=h/2e^2$. This question has given rise to considerable theoretical and experimental works, starting with the contributions of Nigg \textit{et al.} \cite{PRL_97_206804_Nigg, Nanotech_044029_Nigg} followed by Zohar \textit{et al}. \cite{PRB_78_165304_Zohar}. In these works, Coulomb blockade effects inside the mesoscopic capacitor are treated within the Hartree-Fock approximation. As a result, the dot density of states is modified by the Coulomb interaction. Moreover, a Coulomb gap, which equals the charging energy $e^2/C$, appears. Along with the density of states, the quantum capacitance is still periodic. Nevertheless, the periodicity is modified from the single particle level spacing $\Delta$ to the renormalized one $\Delta^{*}$. The value of the charge relaxation resistance is not affected, as it is still quantized to $h/2e^2$, and is therefore  independent of transmission at low temperature. Basically, equations (\ref{EqCqT}) and (\ref{EqRqT}) still hold, but the non-interacting density of states  $N(\epsilon)$ has to be replaced by the one calculated with Coulomb interactions at the Hartee-Fock level. If interactions are not too strong, \textit{i.e.} $e^2/C \approx \Delta$, the change is small and can still be taken into account, by simply replacing $\Delta$ by $\Delta^{\star}$ in the non-interacting calculation. Therefore, our results can be well explained by a simple non-interacting theory. Essentially, the quantization of the charge relaxation resistance has recently been proven to be robust to Coulomb interactions beyond the Hartree-Fock approximation \cite{Nat_Phys_1_690_Mora, PRB_81_153305_Hamamoto}. Regarding the limit of weak and large transparencies of the dot barrier, Mora \textit{et al.} and Hamamoto \textit{et al.} performed analytical calculations \cite{Nat_Phys_1_690_Mora, PRB_81_153305_Hamamoto} and numerical simulations \cite{PRB_81_153305_Hamamoto}, which applied an exact treatment of Coulomb interactions. In their investigations, they have shown that charge relaxation resistance was indeed universal and independent of the interaction strength.

\subsection{Quantum \textit{vs} classical RC circuit}

Kirchhoff's laws prescribe the addition of resistances in series. Its failure has been a central issue in developing our understanding of electronic transport in mesoscopic conductors. Indeed, coherent multiple electronic reflections between scatterers in the conductor were found to make the conductance nonlocal \cite{IBM_1_233_PhilMag_21_863_Landauer}. In the case of the fully coherent quantum RC circuit at gigahertz frequencies, we have shown that a counterintuitive modification of the series resistance led to a situation, in which the resistance is no longer described by the Landauer formula and as such, does not depend directly on transmission. When the resistor transmits a single electronic mode, a constant resistance is found that is equal to half of a resistance quantum, $R_q=h/2e^2$. This resistance, which was modified by the presence of the coherent capacitor, is then termed a ``charge relaxation resistance'', so as to distinguish it from the usual dc resistance, which is wedged between macroscopic reservoirs and described by the Landauer formula. Essentially, it raises important questions regarding the crossover between the fully coherent and fully incoherent mesoscopic capacitor. For instance, in the case of our experimental study, how do we recover the two terminal resistance $(h/e^2)(1/\mathcal{T})$ for the charge relaxation resistance in the fully incoherent case?

At finite temperature, the quantum capacitance and the charge relaxation resistance in equations (\ref{EqCqT}) and (\ref{EqRqT}) have to be thermally averaged to take into account the finite energy width of the electron source so that capacitance oscillations are washed out by thermal broadening and the charge relaxation resistance becomes transmission-dependent. In particular, in the regime $k_BT \gg \Delta$, we find:

\begin{equation}\label{eq:decoherence}
R_q = \frac{h}{2e^2}+ \frac{h}{e^2} \, \frac{1-\mathcal{T}}{\mathcal{T}}
\end{equation}

\noindent which is the series association of a single interface resistance $R_c=h/2e^2$ and the four point QPC resistance given by the Landauer resistance $(h/e^2)(1-\mathcal{T})/\mathcal{T}$ \cite{PRL_31_6207_Buttiker,IBM_1_233_PhilMag_21_863_Landauer,IBM_32_384_Stone}. This means that the QD does not act like an additional reservoir. The thermal broadening seems to act at high temperature as
decoherence although the circuit remains fully coherent. In a more detailed theoretical investigation conducted by Nigg \textit{et  al.} \cite{PRB_77_085312_Nigg}, the loss of coherence in the QD is modeled by attaching a fictitious dephasing or voltage probe thereto. The probe draws no net current and an electron entering the probe is replaced by an electron without any phase correlation. In the case of a dephasing probe, the net current vanishes for each energy. However, it vanishes on average for the voltage probe. The coupling between the QD and the probe gives the strength of the decoherence. As a result, dephasing and voltage probes are indistinguishable for a single channel probe, such that in the fully incoherent regime (coupling at unit transmission between the QD and the probe), it yields the same result as in the thermal broadening regime (see equation  (\ref{eq:decoherence})).

\begin{figure}[h]
\begin{center}
\includegraphics[width=0.6\linewidth]{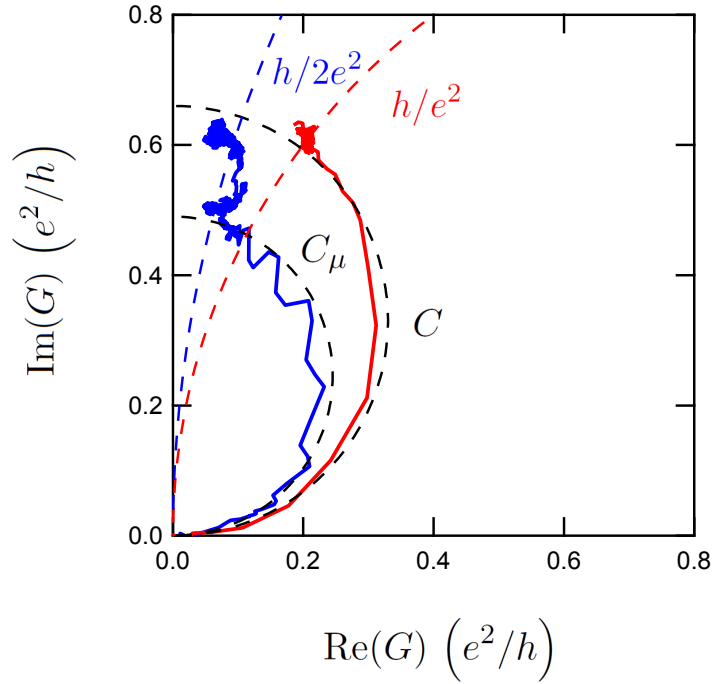}
\end{center}
\caption{Nyquist representation $\mathrm{Re}(G)$ \textit{vs} $\mathrm{Im}(G)$ of the conductance in sample S1 at $B=1.3 \,
\mathrm{T}$ and $\omega/2\pi=1.085 \, \mathrm{GHz}$  in the fully coherent regime at $T=295 \, \mathrm{mK}$ (blue line) and in the high temperature regime at $T=4.2 \, \mathrm{K}$ (red line). Black dashed lines represent Nyquist representations of an RC circuit with a tunable resistance and a constant capacitance ($C_{\mu}(\mathcal{T}=1)$ at $T=295 \, \mathrm{mK}$ and $C$ at $T=4.2 \, \mathrm{K}$). The blue and red dashed lines correspond to Nyquist representations of an RC circuit with a tunable capacitance and a constant resistance $R_q=h/2e^2$ ($R_K=h/e^2$).}
\label{fig10}
\end{figure}

Experimentally, it is difficult to perform a calibration for all temperatures to compare theoretical models with the experiment. Figure \ref{fig10} shows the Nyquist representation $\mathrm{Re}(G)$ \textit{vs} $\mathrm{Im}(G)$ of the RC circuit at low ($T= 295 \, \mathrm{mK}$) and hight ($T= 4.2 \, \mathrm{K}$) temperatures. At low temperature, the circuit is fully coherent and the relaxation resistance is $h/2e^2$ for a single perfectly open channel and draws near to $h/4e^2$ for two open channels. At high temperature when the circuit is expected to be incoherent, the Nyquist representation exhibits a classical behavior of an RC circuit with a tunable resistance and a constant capacitance. For a single perfectly open channel, this resistance approaches the resistance $h/e^2$. It corresponds to the resistance of the QPC at transmission $\mathcal{T}=1$, which indicates that the QD acts like an additional reservoir that can be understood by the presence of a dephasing mechanism involving many coupled channels in the dot.

\section{Experimental setup and calibration}\label{section4}

\subsection{Samples}

The RC mesoscopic circuit is made of the series association of a QPC and a cavity realized in a two-dimensional electron gas (2DEG) in a high-mobility GaAs/GaAlAs heterojunction. We present results on three samples (S1, S3 and S4) measured at low temperatures down to $30 \, \mathrm{mK}$. S1 and S3 have a nominal density $n_e=1.7 \times 10^{11} \, \mathrm{cm}^{–2}$ and mobility $\mu_e=2.6 \times 10^6 \, \mathrm{V}^{-1}.\mathrm{m}^2.\mathrm{s}^{-1}$, while S4 has $n_e=1.9 \times 10^{11} \, \mathrm{cm}^{–2}$ and  $\mu_e=1.3 \times 10^6 \, \mathrm{V}^{-1}.\mathrm{m}^2.\mathrm{s}^{-1}$. S3$^{\star}$, which is actually sample S3, has undergone several thermal cycles resulting in a variation of several parameters like $C_{\mu}$. One of these samples is displayed in figure \ref{fig4}. A finite magnetic field ($B= 1.3
\, \mathrm{T}$) is applied, so as to work in the ballistic integer quantum Hall regime with no spin degeneracy. The mesoscopic cavity has a square shape (see figure \ref{fig4}) and is coupled to a top gate \textit{via} a capacitance $C$. When the QPC is closed, electronic states in the cavity are quantized and a level spacing $\Delta$ can be defined. Transport in the quantum Hall regime is well understood in terms of transport through 1D channel, which allows to evaluate $\Delta$ by using an estimated drift velocity $v_d \sim 5 \times 10^4 \, \mathrm{m}.\mathrm{s}^{-1}$ at $B=1.3 \, \mathrm{T}$ and the size of the cavity, which is deduced from the geometric capacitance $C$ measured at unit transmission (see figure \ref{fig8}). The value of capacitance $C_{\mu}$ is determined independently using the Coulomb blockade spectroscopy in small transmissions, as developed in Section \ref{sec:calibration}. All the characteristics are reported in Table \ref{tab:samples}.

\begin{table}[t]
\caption{\label{tab:samples} Samples characteristics. The capacitance $C$ and the level spacing $\Delta$  are estimated by measuring in unit transmissions ($r\simeq0$), while the level spacing  $\Delta^{\star}$ and the capacitance $C_{\mu}$ are estimated from Coulomb blockade calibrated in small transmissions ($r\simeq1$).}
\vspace{0.5cm}
\begin{indented}
\item[]\begin{tabular}{@{}llllll}
\br
sample &  $C \, (\mathrm{fF})$&  $\Delta \, (\mathrm{GHz})$& $\Delta^{\star} \, (\mathrm{GHz})$& $C_{\mu} \, (\mathrm{fF})$\\
\mr
S1&$\sim8$&$\sim17$&$17\pm2$&$2.3\pm0.3$\\
S3&$\sim4$&$\sim35$&$39\pm4$&$1.0\pm0.07$\\
S3$^{\star}$&$\sim3.5$&$\sim41$&$52\pm4$&$0.75\pm0.07$\\
S4&$<1$&$>60$&$88\pm4$&$0.44\pm0.03$\\
\br
\end{tabular}
\end{indented}
\end{table}

\subsection{Measurement of complex conductance in the microwave regime}

\begin{figure}[h]
\begin{center}
\includegraphics[width=0.8\linewidth]{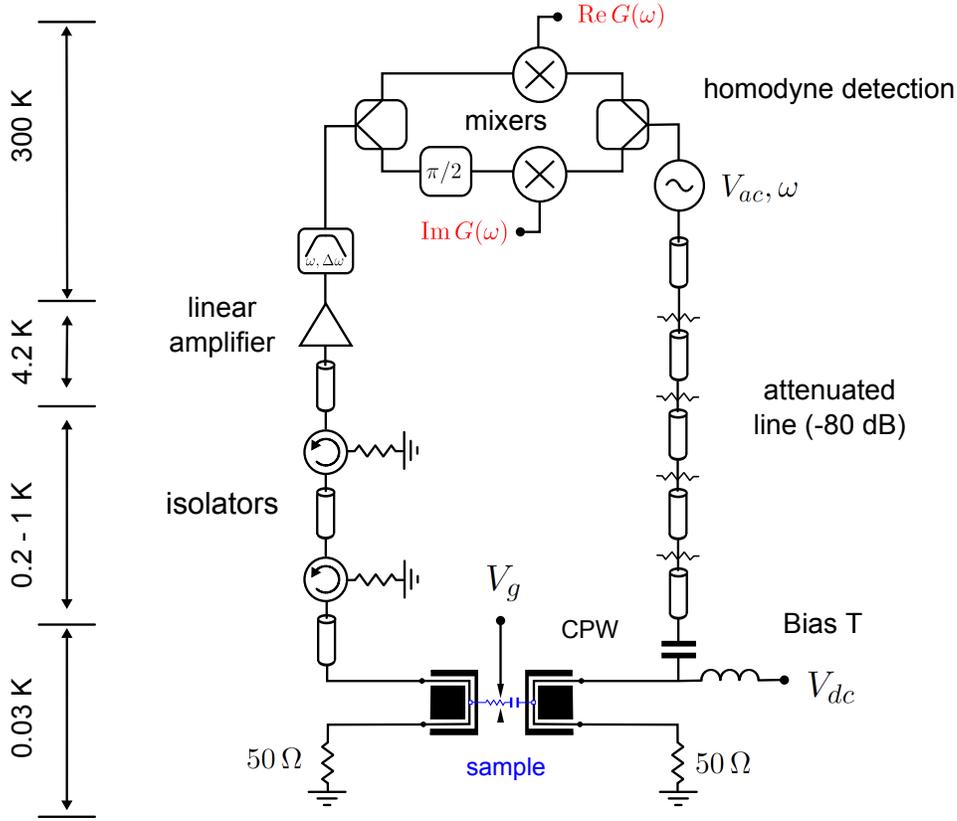}
\end{center}
\caption{Experimental setup for the measurement of the complex conductance $G(\omega)$  of the mesoscopic RC circuit.}
\label{fig11}
\end{figure}

The experimental setup is represented in figure \ref{fig11}. The mesoscopic circuit is cooled down to $30 \, \mathrm{mK}$ in a dilution refrigerator and inserted between two $50 \, \Omega$  coplanar (CPW) transmission lines. The right-hand line is used for excitation, while the left-hand line is used for detection. The excitation $V_{ac} \cos \omega t$ is applied on the top gate of the mesoscopic RC circuit while the response current $I_{ac}
\cos(\omega t + \varphi)$ is measured on the load resistor $R_0=50 \, \Omega$ of the detection line. The principle of the measurement is based on the homodyne technique, where the detected signal is multiplied, after amplification, with the reference signal to provide access to the in-phase response $I_{ac} \cos{\varphi}$ and the out-of-phase response $I_{ac} \sin \varphi$ of the coherent circuit. Note that the impedance of the coherent circuit is on the order of magnitude of the resistance quantum, meaning that the perturbation induced by the circuit between the CPW lines is insignificant and allows for broadband measurement between $0.1$ and $2 \, \mathrm{GHz}$.

Essentially, our goal is to measure the linear response of the coherent RC circuit when the first spin-polarized channel is opened. According to Section \ref{section2}, this regime requires $eV_{ac} \ll \hbar \omega \ll \Delta$ \footnote{More precisely, the condition refers to the potential in the cavity: $eU \ll \hbar \omega \ll \Delta$. According to section \ref{section2}, $U/V_{ac}=\sqrt{\frac{1+1/(R_qC_q\omega)^2}{1+1/(R_qC_{\mu}\omega)^2}}
= \sqrt{\frac{1+1/(\tau_d\omega)^2}{1+1/(\tau_{RC}\omega)^2}}\sim 1$.}. Thus, the experimental setup aims at constraining the frequency of the excitation (i) and its magnitude (ii). Regarding (i), the estimated level spacing in the mesoscopic capacitance $\Delta \sim 15 \, \mathrm{GHz}$  gives an upper limit of the working frequency. The lower limit will be set by the sensitivity we need to measure the RC-time $\tau_{RC}=R_q \times
C_{\mu}$. It can be estimated from the value of the geometric capacitance: $C \sim 5 \, \mathrm{fF}$. Thus, a measuring frequency in the range of $0.1 \sim 2 \, \mathrm{GHz}$  will give the necessary phase sensitivity to measure $\varphi = \arctan \left(\omega \tau_{RC}\right) \sim 0.1$. Concerning (ii), at $\omega /2\pi = 1 \, \mathrm{GHz}$, the linear regime is reached for excitation smaller than $4 \, \mu \mathrm{V}$. Then, it is necessary to compare the magnitude of the excitation with the voltage fluctuations due to thermal photons being brought back by the broadband microwave lines. For a bandwidth $\Delta f \sim 2 \, \mathrm{GHz}$, the thermal noise at room temperature is estimated at $\Delta V_N = \sqrt{4k_BTR_0 \Delta f} \sim 40 \, \mu \mathrm{V}$. The excitation line is thus steadily attenuated by a set of attenuators reducing the thermal fluctuations to $\Delta V_N \sim 0.9 \, \mu \mathrm{V}$ after a total attenuation of $- 80 \, \mathrm{dB}$. Similarly, the coherent conductor has to be isolated from the noise coming from the amplifier. Since attenuators cannot be used without signal loss, cryogenic isolators are therefore used and provide $-30 \, \mathrm{dB}$  of isolation from radiation emitted by the amplifier in the detection chain.

\subsection{Calibration\label{sec:calibration}}

\begin{figure}[h]
\begin{center}
\includegraphics[width=0.75\linewidth]{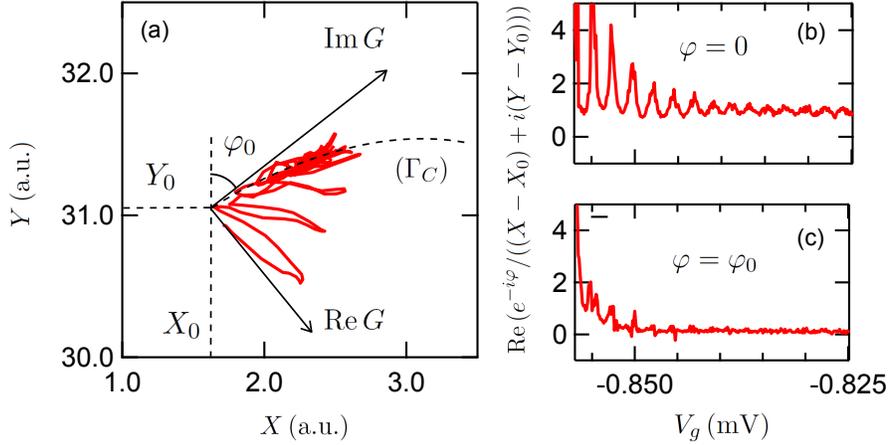}
\end{center}
\caption{(a) Nyquist representation $Y$ \textit{vs} $X$ of raw data performed on sample S2 for $T= 100 \, \mathrm{mK}$, $B=1.3
\, \mathrm{T}$ and $\omega/2 \pi= 1.2 \, \mathrm{GHz}$. Black dashed lines represent one part of the circle $(\Gamma_C)$ corresponding to the Nyquist representation of an RC circuit with a constant resistance and a varying capacitance. (b) and (c) Real part of the complex impedance $\mathrm{Re}
\,(e^{-i\varphi}/((X-X_0)+i(Y-Y_0)))$ numerically computed for two different phase references $\varphi=0$ and $\varphi=\varphi_0$.}
\label{fig12}
\end{figure}

Several quantities need to be determined in order to measure the absolute value of charge relaxation resistance. The raw in-phase and out-of-phase signals given by the homodyne detection can be written as follows:

\begin{eqnarray}
X=X_0+|G| I(\omega) \cos(\varphi-\varphi_0)\\
Y=Y_0+|G| I(\omega) \sin(\varphi-\varphi_0)
\end{eqnarray}

\noindent where $G=|G|e^{i\varphi}$ is the complex conductance of the sample and $I(\omega)$ the oscillating current imposed to the sample. Calibration of the experimental setup thus requires a background subtraction $(X_0,Y_0)$ and a global phase rotation $\varphi_0$. Figure \ref{fig12}(a) shows the Nyquist representation of $Y$ \textit{vs} $X$  at the opening of the first conductance channel. The background $(X_0,Y_0)$ corresponds to a closed QPC (pinched state). It is attributed to the cross-talk between the two CPW transmission lines and can be easily subtracted. Note that its magnitude is approximately $30$ times greater than the RC circuit signal in agreement with an
isolation of $-20 \, \mathrm{dB}$ between the two lines. Determining the global phase  $\varphi_0$ is a more complicated issue that requires using the characteristics of the coherent circuit. Indeed, at the opening of the QPC, the Nyquist representation resembles a fingerprint of the mesoscopic RC circuit. As can be seen from figure \ref{fig6}, starting from the pinched state, peaks are observed in both $\mathrm{Re}(G)$ and $\mathrm{Im}(G)$, but those in $\mathrm{Re}(G)$ quickly disappear while $\mathrm{Im}(G)$ oscillates around a plateau. In the Nyquist representation of figure \ref{fig12} (a), these data points fall on a circle centered on the $\mathrm{Re}(G)$ axis, which characterizes a transmission independent resistance ($(\Gamma_C)$ on figure \ref{fig12} (a)). More precisely, $\varphi_0$ is adjusted to minimize the oscillations of the real part of the complex impedance. Figures \ref{fig12}(b) and (c) show the real part of the complex impedance after numerical inversion with and without phase reference correction.

\begin{figure}[h]
\begin{center}
\includegraphics[width=0.75\linewidth]{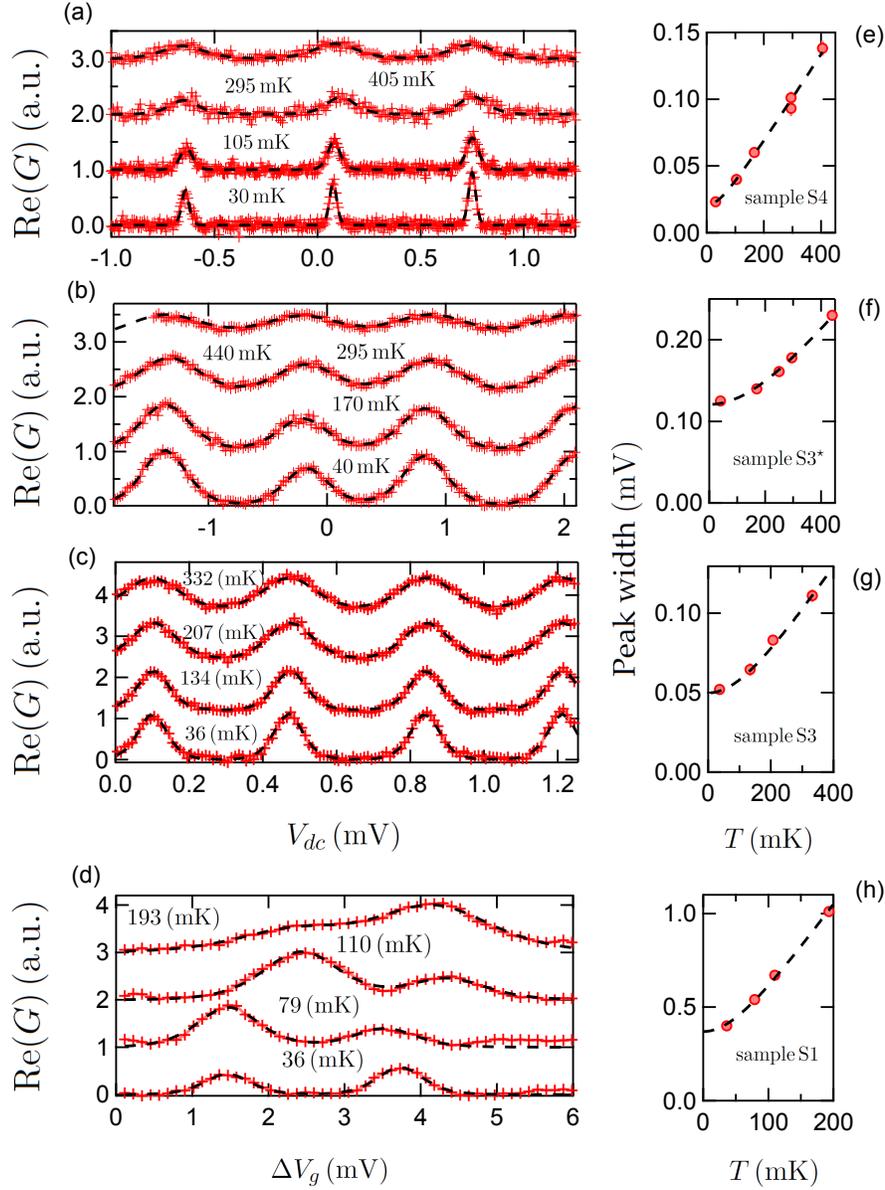}
\end{center}
\caption{Coulomb-blockade oscillations in the real part of the ac conductance in the low-transmission regime. The control voltage is applied to the counter-electrode for sample S2 (b) and to the QPC gate for sample S1 (a). The temperature dependence is used for the absolute calibration of our setup, as described in the text, where the peak width, shown in (c) and (d) as a function of temperature, is deduced from theoretical fits (dashed lines) using equation (\ref{EqRqbraod}), and taking a linear dependence of energy with the control voltage. Lines in (c) and (d) are fits of the experimental results derived from using a $\sqrt{T_{el}^2+T^2}$ law to take into account a finite residual electronic temperature  $T_{el}$.}
\label{fig13}
\end{figure}

The last step of the calibration procedure requires calibrating the whole detection chain. However, at GHz frequencies, direct calibration is hardly better than at $3 \, \mathrm{dB}$ . For this purpose, we will use an indirect, but absolute method often used in Coulomb blockade spectroscopy. The method is based on comparing the gate voltage width of a thermally broadened Coulomb peak ($\sim k_BT$) and the Coulomb peak spacing ($\sim e^2/C_{\mu}$). As a result, an absolute value of $C_{\mu}$  can be obtained. The real part of the admittance in samples S3, S3$^{\star}$ and S4 is shown as a function of the dc voltage $V_{dc}$ at the counter-electrode for a given low transmission (see figure \ref{fig13} (a)(b)(c)). A series of peaks with periodicity $\Delta V_{dc}$  are observed, with the peaks accurately fitted by using equation (\ref{EqRqbraod}). Their width, which is proportional to the electron temperature $T_{el}$, is plotted against the refrigerator temperature $T$ (see figure  \ref{fig13} (e)(f)(g)). When corrected for apparent electron heating arising from gaussian environmental charge noise, i.e. $T_{el}=\sqrt{T_0^2+T^2}$, the energy calibration of the gate voltage yields $C_{\mu}$ and the amplitude $1/C_{\mu}\omega$ of the conductance plateau in figure \ref{fig6}. A similar analysis was performed in figures \ref{fig13} (d) and (h), for sample S1 using $V_{g}$ to control the cavity potential. Here, peaks are distorted due to a transmission-dependent background and show a larger periodicity $\Delta V_g= 0.2 \, \mathrm{mV}$, which reflects the weaker electrostatic coupling to the 2DEG

Looking at the samples characteristics (see table \ref{tab:samples}), we noticed that the estimated level spacing $\Delta$ (related to the cavity size) and the measured one $\Delta^{\star}=e^2/C_{\mu}$ differs from a factor of $\sim 1.2$ for the different samples. From the data gathered, we observed that the smaller cavity is, the greater the difference is. This difference is related to the charging energy $e^2/C$, which corresponds to the energetic cost (Coulomb interactions) associated to the addition of one electron on the capacitor $C$. As discussed in Section \ref{Coulomb}, this effect can be considered when replacing the bare single particle level spacing $\Delta $ by the addition energy: $\Delta^{\star}= \Delta + e^2/C$.

\section{Conclusion}

To investigate the effect of quantum coherence on electronic dynamics in quantum conductors, we studied a model quantum conductor, \textit{i.e.} the quantum RC circuit, which comprises a single channel spin polarized electronic cavity that is capacitively coupled to a metallic top gate and tunnel coupled to an electronic reservoir by a quantum point contact of tunable transmission. This circuit realizes the quantum version of the well-known  RC circuit in electronics, where the dynamics of charge transfer are encoded in the charge relaxation time $\tau_{RC}=RC$. In a quantum conductor, where the phase coherence is preserved along the electronic path, one cannot analyze the various components of the conductor individually. Consequently, in the model circuit studied in this review, the charge relaxation time cannot be understood merely from the serial association of the resistance of the quantum point contact $R=h/(e^2\mathcal{T})$ measured in dc transport measurements and the geometric capacitance $C$  of the cavity. Rather, due to quantum interferences, the circuit has to be considered as a whole entity. Although charge relaxation is still analogous to that of an RC circuit, the capacitance is given by the serial association of the geometric capacitance $C$ and a quantum capacitance $C_q$ related to the density of states in the cavity while the charge relaxation resistance is quantized to $R_q=h/2e^2$ and independent of the QPC transmission $\mathcal{T}$. This remarkable manifestation of quantum coherence in dynamic transport, which was first theoretically predicted by M. B\"{u}ttiker, H. Thomas and A. Pr\^{e}tre \cite{Phys_Lett_A180_364_Buttiker}, is confirmed by our experimental study. Moreover, the detailed behavior of this model circuit can be verified with great accuracy within the context of the quantum scattering theory of dynamic transport.

In recent years, many interesting developments in this topic have emerged. However, many questions remain open. For instance, the case of metallic boxes with many conducting channels and a small dot level spacing has been theoretically investigated in \cite{PRL_106_166803_Ledoussal}, where the authors show that a certain average of the charge relaxation resistance is still being quantized. On the experimental side, other types of quantum conductors have been considered, such as carbon nanotubes in single dot \cite{PRL_107_256804_Delbecq} or double dot geometries \cite{PRL_108_036802_Chorley}. The latter has a significant impact in quantum information applications, where the mesoscopic admittance of the double dot plays a crucial role \cite{PRB_83_121311_Cottet,PRL_108_036802_Chorley}. The former addresses the fundamental question of the dynamic properties of a quantum conductor in a different regime, where the electronic correlations are strong, namely the Kondo regime. Carbon nanotubes represent a distinct type of one-dimensional conductors compared to the edge channels of the quantum Hall regime. Two orbital channels carrying two types of spin species participate in the transport. When the spin degeneracy is taken into account, one should distinguish the charge dynamics characterized by the charge susceptibility from the spin dynamics related to the spin susceptibility. In the Kondo regime, a single spin is trapped inside the dot and behaves like the magnetic impurity coupled to the electrons in the leads of the original Kondo problem. Due to Coulomb interactions, when the charge inside the dot is frozen, the charge susceptibility is small. However, the spin can fluctuate due to the coupling to the electrons in the conductor leads. These spin fluctuations lead to a resonant peak in the density of states at the Fermi energy, which can be observed through dc conductance measurements of the dot \cite{Nature_391_156_Goldhaber, Science_289_2105_Wiel}. However, recent theoretical works \cite{PRB_83_201304_Lee,PRL_107_176601_Filippone} have predicted that the capacitance remains small in this regime, since it is related to charge susceptibility. In this example and contrary to the case considered in this review, where Kondo correlations were absent, the capacitance should differ from the density of states. Moreover, if the spin degeneracy were lifted by applying a magnetic field, the Kondo state would be destroyed, and a strong
increase of the charge relaxation resistance would be expected for
intermediate magnetic fields. In this case, the charge relaxation resistance would strongly deviate from its quantized value. Another case where electronic correlations play a crucial role is the fractional quantum Hall effect. It has been predicted \cite{PRB_81_153305_Hamamoto} that the universality of the charge relaxation resistance $R_q=h/(2\nu e^2)$, where $\nu<1$ is the fractional filling factor, would break down for values of $\nu <1/2$, thereby also resulting in a divergence of the charge relaxation resistance.

Finally, another case of interest is the dynamics of charge
transfer when the conductor is put out of equilibrium. In this case, the current generated by a voltage excitation of the dot should be computed beyond the linear response considered in this review. In this regime, the mesoscopic cavity can act as a single electron source that emits a quantized number of charge \cite{Science_316_1169_Feve, PRL_100_086601_Moskalets} in the conductor. Also here, in the presence of Coulomb interactions, the interplay between the charge and spin degrees of freedom has to be considered in order to understand the charge and spin dynamics of dot \cite{Science_316_1169_Feve, PRL_100_086601_Moskalets}. These issues are of prime importance when manipulating single spins or single charges in quantum conductors.

\section*{Acknowledgments}

We are very grateful to D.C Glattli and Y. Jin for their major contribution to the work presented here. We thank M. B\"{u}ttiker and S. Nigg for numerous insightful discussions on our experimental investigations and the keen interest and encouragement they have shown in our endeavors. The author would like to express his sincere thanks to Joel Moser for his careful reading of the manuscript.

\end{document}